\documentclass[10pt, a4paper, preprint]{elsarticle}

\journal{Information Systems}

\usepackage{fullpage}
\usepackage{comment}
\usepackage{url}
\usepackage{colortbl}
\usepackage[usenames]{xcolor}
\usepackage{epstopdf}              
\usepackage{hologo}
\usepackage{multirow}
\newcommand{\uilib}{{\tt uiHRDC}}

\newcommand{\svs}{\textit{svs}}

\newcommand{\merge}{\textit{merge}}

\newcommand{\repair}{Re-Pair}
\newcommand{\lzend}{LZ-End}
\newcommand{\lzma}{LZMA}

\newcommand{\CM}{\texttt{CM}}
\newcommand{\ST}{\texttt{ST}}

\newcommand{\vbyte}{\texttt{Vbyte}}
\newcommand{\vbyteB}{\texttt{VbyteB}}
\newcommand{\rice}{\texttt{Rice}}
\newcommand{\riceB}{\texttt{RiceB}}
\newcommand{\simplen}{\texttt{Simple9}}

\newcommand{\pfordelta}{\texttt{PforDelta}}
\newcommand{\qmx}{\texttt{QMX}}
\newcommand{\riceRuns}{\texttt{Rice-Runs}}

\newcommand{\vbyteCM}{\texttt{Vbyte-CM}}
\newcommand{\vbyteCMB}{\texttt{Vbyte-CMB}}
\newcommand{\vbyteST}{\texttt{Vbyte-ST}}
\newcommand{\vbyteSTB}{\texttt{Vbyte-STB}}
\newcommand{\repairNo}{\texttt{RePair}}
\newcommand{\repairSkip}{\texttt{RePair-Skip}}
\newcommand{\repairSkipCM}{\texttt{RePair-Skip-CM}}
\newcommand{\repairSkipST}{\texttt{RePair-Skip-ST}}
\newcommand{\vbyteLZMA}{\texttt{Vbyte-LZMA}}
\newcommand{\vbyteLzend}{\texttt{Vbyte-Lzend}}
\newcommand{\rlcsa}{\texttt{RLCSA}}
\newcommand{\wcsa}{\texttt{WCSA}}
\newcommand{\slp}{\texttt{SLP}}
\newcommand{\wslp}{\texttt{WSLP}}
\newcommand{\lzindex}{\texttt{LZ77-index}}
\newcommand{\lzendindex}{\texttt{LZend-index}}

\newcommand{\interpolative}{\texttt{Interpolative}}
\newcommand{\efopt}{\texttt{EF-opt}}
\newcommand{\optpfd}{\texttt{OPT-PFD}}
\newcommand{\varint}{\texttt{varintG8IU}}

\begin{document} 
\begin{frontmatter}	
	\title{On the Reproducibility of Experiments of Indexing Repetitive Document Collections}

	\author[udc]{Antonio Fari\~na\corref{cor1}}
	\ead{antonio.farina@udc.es}
		
	\author[uva]{Miguel A. Mart\'inez-Prieto}
	\ead{migumar2@infor.uva.es}
		
	\author[udp]{Francisco Claude}
	\ead{fclaude@recoded.cl}
		
	\author[uchile]{Gonzalo Navarro}
	\ead{gnavarro@dcc.uchile.cl}
			
	\author[uned]{Juan J. Lastra-D\'iaz\corref{rr}}
	\ead{jlastra@invi.uned.es}
			
	\author[pisa]{Nicola Prezza\corref{rr}}
	\ead{nicola.prezza@gmail.com}
			
	\author[udconce]{Diego Seco\corref{rr}}
	\ead{dseco@udec.cl }

	\cortext[cor1]{Corresponding author}	
	\cortext[rr]{Reproducibility reviewer.}
	
	\address[udc]{University of A Coru\~na, Facultade de Inform\'atica, CITIC, Spain.\\}
	\address[uva]{DataWeb Research, Department of Computer Science, University of Valladolid, Spain.}		
	\address[udp]{Escuela de Inform\'atica y Telecomunicaciones, Universidad Diego Portales, Chile.\\}
	\address[uchile]{Millennium Institute for Foundational  Research on Data (IMFD), Department of Computer Science, University of Chile, Chile.}
	\address[uned] {Universidad Nacional de Educaci\'on a Distancia, Spain.}
	\address[pisa] {University of Pisa, Italy.}
	\address[udconce]{Millennium Institute for Foundational  Research on Data (IMFD),Department of Computer Science, Faculty of Engineering, University of Concepci\'on, Chile.}

	\begin{abstract}
	This work introduces a companion reproducible paper with the aim of allowing the exact replication of the methods, experiments, and results discussed in a previous work \cite{CFMPN:16}. In that parent paper, we proposed many and varied techniques for compressing indexes which exploit that highly repetitive collections are formed mostly of documents that are near-copies 
	of others. 
	More concretely, we describe a replication framework, called \uilib\ {\em (universal indexes for 
	Highly Repetitive Document Collections)}, that allows our original experimental setup to be easily replicated 
	using various document collections. The corresponding experimentation is carefully explained, providing 
	precise details about the parameters that can be tuned for each indexing solution. Finally, note that we also 
	provide \uilib\ as reproducibility package.  

	\bigskip
	\noindent

	\end{abstract}

\end{frontmatter}

\section{Introduction}
\label{sec:1}
Scientific publications remain the most common way for disseminating scientific advances. Focusing on Computer
Science, a scientific publication: {\em paper}, i) exposes new algorithms or techniques for addressing an 
open challenge, and ii) reports experimental numbers to evaluate these new approaches regarding a particular 
baseline. Thus, empirical evaluation is the stronger evidence of the achievements reported by a paper, and
its corresponding setup is the only way that the scientific community has for assessing and (possibly) 
reusing such research proposals. 

The ideal situation arises when a paper exposes enough information to make its research easily reproducible. 
According to the definition proposed by the Association for Computation Machinery (ACM) \cite{ACM:18}, an 
experiment is {\em reproducible} when an independent group of researchers can obtain the same results using 
artifacts which they develop independently. A weaker form of reproducibility is replicability. In this case,
the original setup is reused, so an experiment is {\em replicable} when an independent group of researchers 
can obtain the same results using the author own artifacts. Finally, the ACM document \cite{ACM:18} also
defines the concept of repeatability. An experiment is {\em repeatable} if the group of researchers is able
to obtain the same results reusing their original setup (including the same measurement procedures and the 
same system, under the same operating conditions). Non-repeatable results do not provide solid insights about 
any scientific question, so they are not suitable for publication. Thus, we can accept that research in
Computer Science must be, at least, repeatable.

Although reproducibility is the ultimate goal, getting it is not always easy because it requires that a new 
experimental setup to be reconstructed from scratch. Implementing new algorithms, data structures, or other 
computational artifacts can be extremely complex, and tuning them may require setting many parameters which, 
in turn, can depend on particular system configurations or external tools. Besides, it may be necessary to 
use programs in the exact versions used originally \cite{Sandve:13}. As a consequence, reproducibility is 
time-consuming, error-prone, and sometimes just infeasible \cite{Chirigati:16}. 

In practice, publishing replicable research\footnote{In this paper, we use the term reproducibility to refer 
experimental setups that can be replicated.} would be an important step forward in Computer Science. According 
to Collberg et al. \cite{Collberg:15}, replicating a computational experiment {\em only} needs that the source 
code and test case data have to be available. Although simple, many papers do not meet this precondition. The
aforementioned paper \cite{Collberg:15} provides an interesting study involving 601 papers from conferences and 
journals. It shows that only $32.3\%$ of the papers provide enough information and resources to build 
their executables in less than 30 minutes. Note that external dependencies must be accessible to compile these
sources. Regarding test case data, Vandewalle et al. \cite{Vandewalle:09} report that more papers make data
available, but it is mainly due to some of them use standard corpora in their corresponding experimental setups.
However, the 
problem goes beyond. Once compiled, proper parameter settings must be set, and sometimes it is tricky to find this 
information in the papers. In conclusion, reproducibility research is currently challenging.

The situation is not quite different in {\bf Information Retrieval} (IR), the field more related to the research 
proposed in our work. IR is a broad area of Computer Science focused primarily on providing the users with easy access 
to information of their interest \cite{BYRN:11}. As an empirical discipline, advances in IR research are built on experimental 
validation of algorithms and techniques \cite{Lin2016}, but reproducing IR experiments is extremely challenging, even when 
they are very well documented \cite{Ferro:15}. A recent Dagstuhl Seminar about {\em reproducibility of data-oriented experiments 
in e-Science} \cite{Ferro:16} concludes that IR systems are complex and depend on many external resources that cannot be 
encapsulated with the system. Besides, it reports that many data collections are private, and their size could be an
obstacle even for obtaining them. Finally, it notes that some experimental assumptions are often hidden, making reproducibility 
a less achievable goal.

Following the initiative of some previous papers \cite{repro1,repro2}, the aim of this work is to propose a detailed 
experimental setup that replicates the methods, experiments, and 
results on {\em indexing repetitive document collections} discussed in a previous work \cite{CFMPN:16} (we will refer
it as the {\em parent paper}). This 
experimental setup focuses on two dimensions: i) the {\em space} used to preserve and manage the document collection, 
and the ii) {\em time} needed to query this data. 

The rest of the paper is organized as follows. In Section \ref{sec:2}, we briefly describe all the inverted-index based 
variants and the self-indexes evaluated in the parent paper. We also explain the most relevant space/time 
tradeoffs reported from our experiments. The following sections are devoted to the reproducibility of these experiments. 
Section \ref{sec:3} details the fundamentals of \uilib, our replication framework, which comprises {\em i)} the {\em source 
code} of all our indexing approaches, {\em ii)} some {\em test data} collections, and {\em iii)} the {\em set of scripts} 
that runs the main tasks of our experimental setup and generates a final report with the main figures of the parent paper. 
We focus both on discussing the workflow that leads the process of 
replicating all our experiments, and also in describing the structure of the \uilib\ framework. Section \ref{sec:docker} 
describes the Docker\footnote{\tt https://www.docker.com/} container that allows this workflow to be easily reproduced in 
a tuned environment. Some brief conclusions are exposed in Section \ref{s:conc} to motivate the need of improving research
reproducibility in Computer Science. Finally, Appendix \ref{app:A} includes the actual compression ratios obtained for each technique (which complements the results shown in the figures throughout the paper), and Appendix \ref{app:B} is devoted to explain how some of our self-indexes can
be reused for dealing with universal (not document oriented) repetitive collections.


\section{Universal Indexes for Highly Repetitive Document Collections}
\label{sec:2}
\subsection{Background}
Indexing highly repetitive collections has gained relevance because huge corpora of versioned documents are 
increasingly consolidated. {\em Wikipedia}\footnote{\tt https://www.wikipedia.org/} is the most recognizable example,
exposing millions of articles which evolve to provide the best picture of the reality around us. Each Wikipedia
article comprises one version per document edition, and most of them are near-duplicates of others. Apart
from versioned document collections, other applications perform on versioned data. For instance, the
version control system {\em GitHub}\footnote{\tt https://www.github.com/}. In this case, a tree of
versions is maintained to control changes from millions of software projects which are continuously updated by 
their developers. Biological databases (where many DNA or protein sequences from organisms of the
same or related species are maintained), or periodic technical publications (where the same data, with small
updates, are published over and over) are other examples of computer systems managing highly repetitive
data.

The parent paper \cite{CFMPN:16} focused on natural language text collections, which can be decomposed into 
words, and queried for words or phrases. Managing these document collections is a challenge by itself due to their 
large volume, but at the same time, they are highly compressible due to their repetitiveness. In addition, 
version control systems require direct access to individual versions. This is also challenging because it demands 
efficient and potentially large indexes to be deployed on top of the data collection. Thus, we need to 
compress not only the data, but also indexes required to speed up searches.

The {\bf Inverted Index} \cite{BYRN:11} has been traditionally used to index natural language text collections.
The inverted index maintains a list of the occurrences (also referred  as {\em posting list} or {\em inverted list}) 
of each distinct word in the text. It enables two different types of inverted indexes to be distinguished: 
i) {\em non-positional} indexes store the list of document identifiers that contain each different word; and ii) 
{\em positional} indexes store, in addition to the document identifiers, the word offsets of the occurrences within 
each document.
Posting lists are usually sorted by increasing document identifier, and by increasing word offset within each 
document for positional indexes. This decision is useful for list {\em intersections}, a fundamental task under 
the Google-like policy of treating bag-of-word queries as ranked AND-queries. Intersections are also relevant
to solve queries involving multiple words (phrase queries).

There is a burst of recent activity in exploiting repetitiveness at the indexing structures, in order to
provide fast searches in the collection within little space. These approaches are summarized in the
following. On the one hand, Section \ref{sec:compII} discusses the fundamentals of inverted index compression, 
and summarizes all our approaches. On the other hand, Section \ref{sec:self} introduces the concept of self-index 
\cite{Nav:16}, an innovative succinct data structure which integrates data and index into a single compressed 
representation. Besides, we explain how self-indexes can be adapted to perform on document versioned collections 
attending to our original work. A detailed review of related literature, and a full explanation of our 
compressed inverted indexes and self-indexes can be found in the parent paper.

\subsubsection{Compressed Inverted Indexes} \label{sec:compII}

Since posting lists are sequences of increasing numbers, traditional compression techniques typically 
compute the difference between consecutive values and then encode those {\em gaps} with a variable-length
encoding that favors small numbers. This is the basis of techniques traditionally used to represent posting lists: 
those using \rice\ codes or \vbyte\ codes, that assign one codeword to each {\em gap}, or others such as
{\simplen}, that packs several {\em gaps} within a unique integer, or \pfordelta\ which compresses blocks of $k$ {\em gaps}. 
Therefore, all aforementioned gap encoding techniques take advantage of the fact that the values within 
posting lists (document identifiers or position values) are probably very close, and consequently they require few bits for their encoding. We have revisited all the previous techniques and we used them in this repetitive scenario. In addition, we have considered other state-of-the-art posting list representations including: \qmx, which is a good representative of the last generation of SIMD-based techniques \cite{trotman2014}, or the recent Elias-Fano technique from \cite{OV14} (\efopt) and the implementations from \cite{OV14} of some well-known representations such as \optpfd\ \cite{YDS09}, \interpolative\ coding \cite{MS00}, and \varint\ \cite{Stepanov:2011}.

Unfortunately, traditional techniques are not able to detect the repetitiveness that arises in versioned collections.
As one of the major contributions of our parent paper, we proposed some different techniques which exploit the types of repetitiveness that arise in versioned collections. We applied them both for non-positional
and positional inverted indexes. They are based on run-length encoding (\riceRuns), grammar-based compression (\repairNo), and Lempel-Ziv compression (\vbyteLZMA\ and \vbyteLzend).

Even though the use of compressed posting list representations permits to drastically reduce inverted index size, it
also has implications in query time. As we indicated in the parent paper, 
intersections can be performed by traversing the lists sequentially in a {\em merge}-type fashion.
However, if one of the lists to intersect is much shorter than the other/s it is preferred to provide direct access 
(using sampling) so that
the elements of the smallest list can be searched for within the longer list. This type of intersection is commonly
referred to as {\em Set-vs-Set} (\svs) in the literature. In practice, the best choice is to sort 
the lists by length. Then, we take the shortest one as the ``candidate'' list, and it is iteratively intersected with
longer and longer lists, hence shortening the candidate list at each step.

There are two main sampling structures to provide direct access in the literature. Culpepper and Moffat \cite{CM10}
propose the first one, referred to as \CM. It regularly samples the compressed list and stores separately an array 
of samples, which is searched with exponential search. 
Given a sampling parameter $k$, a list of length $\ell$ is sampled every $k \log_2 \ell$ 
positions. 
The second method, by Transier and Sanders \cite{TS10} (\ST), applies domain sampling. It regularly samples the universe of 
positions, so that the exponential search can be avoided. Given a parameter $B$, 
it samples the universe of size $u$ at intervals $2^{\lceil \log_2 (uB/\ell) \rceil}$. To perform 
intersections,  {\em lookup} algorithm was defined \cite{TS10}. It somehow resembles a {\em merge}-type 
algorithm but takes advantage of the domain sampling. As we will see below, we combined both \CM\ and \ST\ sampling
with posting list representations based on \vbyte\ and with our \repair-based alternatives.



\begin{table}
	\centering
	\footnotesize
	\begin{tabular}{|l|l|p{9cm}|}
		\hline
		{\bf Method}                                 & {\bf Variants} & {\bf Description} \\ 
		\hline
		\multirow{6}{*}{\vbyte}						 & \vbyte   \cite{WZ99}	 &	
				Simple Vbyte encoding with no sampling.	Intersections are performed in a {\em merge-wise} fashion.\\ \cline{2-3}
													 & \vbyteB	\cite{CM10}			 &										 
				\vbyte\ enhanced with bitmaps to represent the longest posting lists. \\ \cline{2-3}
													 & \vbyteCM \cite{CM10}	 &	
				\vbyte\ coupled with list sampling: $k = \{4,32\}$. Intersections are perfomed in a {\em set-vs-set} approach.  \\ \cline{2-3}
													 & \vbyteCMB \cite{CM10} &	
				\vbyteCM\ enhanced with bitmaps to represent the longest posting lists. \\ \cline{2-3}
													 & \vbyteST \cite{TS10}  &	
				\vbyte\ coupled with domain sampling: $B = \{16,128\}$ ($B = \{64\}$ is also tested for the positional scenario). Intersections follow a {\em lookup} approach. \\ \cline{2-3}
													 & \vbyteSTB \cite{CFMPN:16}		 	 &	
				\vbyteST\ enhanced with bitmaps to represent the longest posting lists. \\ \hline
		\multirow{2}{*}{\rice} 						 & \rice \cite{WMB99}	 &	
				Simple Rice encoding with no sampling. Intersections are performed using a {\em merge} algorithm.	\\\cline{2-3}
													 & \riceB \cite{CFMPN:16}				 &	
				\rice\ \cite{WMB99} enhanced with bitmaps to represent the longest posting lists.  \\ \hline
		\simplen\ \cite{AM05}              			 & No variants           &  
				Word-aligned \simplen\ that packs consecutive {\em gaps} into a 32-bit words. Intersections are performed using a {\em merge} algorithm.   \\ \hline
		\pfordelta\ \cite{Hem05,ZHNB06}     		 & No variants           & 
				Extends \simplen\ {\em gaps} to pack many {\em gaps} together (up to 128). A variant of \simplen\ is used to encode exceptions. Intersections are performed in a {\em merge-wise} fashion.      \\ \hline
		\qmx\ \cite{trotman2014, Lemire2015:simdInt} & No variants           & 
				Exploits SIMD-instructions to boost the decoding of long lists. Intersections are performed using an specific algorithm that also benefits from such instructions.     \\ \hline
		\riceRuns\ \cite{CFMPN:16}                   & No variants           & 
				 \rice\ \cite{WMB99} coupled with {\em run-length} encoding. Intersections are performed in a {\em merge-wise} fashion.     \\ \hline
		\vbyteLZMA\ \cite{CFMPN10}                  & No variants           &       
				Encodes {\em gaps} with \vbyte\ \cite{WZ99} and, if the size of the resulting \vbyte-sequence is $\geq$ 10 bytes, then it is further compressed with {\em LZMA}. A bitmap indicates which posting lists are compressed with LZMA. Intersections are performed using  {\em merge} algorithm.  \\ \hline
		\vbyteLzend\ \cite{CFMPN:16}                  & No variants           & 
				Encodes {\em gaps} with \vbyte\ \cite{WZ99} and then all posting lists are compressed as a whole with {\em LZ-End} \cite{KN:12}. It can be parameterized: $ ds =\{4,16,64,256\}$. Intersections first obtain \vbyte-encoded sequences and then perform in a {\em merge-wise} fashion.      \\ \hline
		\repairNo-based \cite{CFMPN:16}              & \repairNo       		 &       
				Compresses  posting lists as a whole using RePair \cite{LM:99}. Intersections are performed using  {\em merge} algorithm over the compressed lists. \\\cline{2-3}
													 & \repairSkip			 &
				Adds {\em skipping} data to \repairNo\ to improve performance.  \\\cline{2-3}
													 & \repairSkipCM		 &  
				
				\repairSkip\ coupled with {\tt CM}-type (list) sampling: $k = \{1,64\}$. \\	\cline{2-3}
													 & \repairSkipST		 &  
				\repairSkip\ coupled with {\tt ST}-type (domain) sampling: $B = \{1024\}$ for the non-positional scenario; $B = \{256\}$  for the positional scenario. \\	\hline
		\multirow{1}{*}{Indexes from Ottaviano  }   & \efopt\ \cite{OV14}   & 					Partitioned Elias-Fano index.		\\\cline{2-3}
		\multirow{1}{*}{and Venturini's }  & \optpfd\ \cite{YDS09} &  				Optimized \pfordelta\ variant.        \\\cline{2-3}
		\multirow{2}{*}{Framework \cite{OV14}}							   & \interpolative\ \cite{MS00}  & 				Binary interpolative coding.   \\\cline{2-3}
													   & \varint\ \cite{Stepanov:2011} &   				SIMD-optimized Variable Byte code. \\
		\hline
	\end{tabular}
	\caption{\label{table} Summary of posting list representations evaluated in the parent paper. We used our own implementations of posting list representations (except those from Ottaviano and Venturini's Framework) that, in most cases, were built over existing techniques to represent integers or general sequences.}	
\end{table}

\paragraph{Posting List Representations}


We include here a brief description of the different posting list representations evaluated in our original paper and the tuning options they support (if any). All this information is summarized in Table \ref{table}.

\begin{itemize}
	\item \vbyte. We included a simple posting list representation based on \vbyte\ \cite{WZ99} which uses no sampling and consequently performs intersections in a {\em merge}-wise fashion.  We also included two alternatives using \vbyte\ coupled with sampling \cite{CM10} (called \vbyteCM), with $k=\{4,32\}$, or domain sampling \cite{TS10} (called \vbyteST), with $B=\{16,128\}$. In the former case, as indicated above, intersections follow a {\em Set-vs-Set approach} where the smallest list is decompressed and its values are searched for within the other lists using exponential search. For \vbyteST\ we used {\em lookup} intersection algorithm \cite{TS10}.
	In addition, we included a hybrid variant of \vbyteCM\ that uses bitmaps to represent the longest posting lists (\vbyteCMB) \cite{CM10}.
	In practice, lists longer than $u/lenBitmapDiv$ (where $u$ is the largest document identifier, and $lenBitmapDiv=8$) are replaced by a bitmap \cite{CM10} that marks which
	documents are present in the list.  For completeness, we used the hybrid approach over \vbyteST\ to build \vbyteSTB, and also included a variant \vbyteB\ with no sampling.
		
	\item \rice. We included a representation based on \rice\ codes \cite{WMB99} and also implemented for completeness a \riceB\ variant using bitmaps for the longest lists. In both cases they use no sampling and intersections are done using {\em merge} algorithm.
	
	\item \simplen. We included a representation based on the word-aligned \simplen\ technique, by Anh and Moffat \cite{AM05}. It packs consecutive {\em gaps} into a $32$-bit word. It uses the first $4$ bits to signal the type of packing done, depending on how many bits the next 
	{\em gaps} need: we can pack $28$ $1$-bit numbers, or $14$ $2$-bit numbers, and so on. 	 Intersections are done using {\em merge} algorithm.
	
	\item \pfordelta. This representation \cite{Hem05,ZHNB06} uses the same idea of packing many {\em gaps} together, typically up to $128$ (parameter $pfdThreshold=128$), while allowing for 10\% of exceptions that need more bits than the core 90\% of the {\em gaps}. Those exceptions are then coded separately using a variant of \simplen. In our case, we obtained our best results with $pfdThreshold=100$. Again, we performed intersections {\merge}-wise.
	
	\item \qmx. This technique \cite{trotman2014} uses SIMD-instructions to boost decoding of large lists,\footnote{\url{http://www.cs.otago.ac.nz/homepages/andrew/papers/QMX.zip}. } and was coupled with an intersection algorithm \cite{Lemire2015:simdInt} that also benefits from SIMD-instructions.\footnote{ \url{https://github.com/lemire/SIMDCompressionAndIntersection}.}
	
	\item \riceRuns. We coupled {\em run-length} encoding with \rice\ coding to boost both the compression and intersection speed of our \rice\ representation in a repetitive scenario where large sequences of $+1$ {\em gaps} could occur. Basically, a sequence of $+1$ {\em gaps} of length $k$ is represented as the {\em ricecode($+1$)} followed by {\em ricecode($k$)}. Intersections are performed {\em merge}-wise, 
	yet they take advantage of the fact that a run of $+1$ values can be skipped/decoded in a unique operation.
	
	\item \vbyteLZMA. In this representation we independently compress each posting list with a \vbyte+{\em \lzma} chain.
	We initially encode the {\em gaps} in the posting list with \vbyte\ and then compress the output
	with the {\em \lzma} variant of LZ77 ({\tt www.7-zip.org}). We use a threshold parameter ($minbcssize$) so that 
	\lzma\ is only applied  on those lists where their \vbyte\ encoded sequence occupies at least $10$ bytes ($minbcssize\leftarrow 10$). Otherwise, the initial \vbyte\ encoded sequence is stored. A bitmap indicates which posting 
	lists were stored compressed with \vbyte\ plus \lzma, and which ones only with \vbyte. Since \vbyteLZMA\ only supports extracting a list from 
	the beginning, intersections can involve a initial step that applies \lzma\ decoder to recover the \vbyte\ encoded sequences, and then we apply {\em merge} algorithm on those \vbyte\ encoded sequences.
	
	\item \vbyteLzend. This representation follows the same idea of \vbyteLZMA\ but uses \lzend\ \cite{KN:12} instead of \lzma\ to compress
	the posting lists. Since \lzend\ allows random access, the sequence of all the concatenated lists can be compressed as a whole, not list-wise. 
	Therefore, we first concatenate the \vbyte\ representation of all the posting lists (keeping track of the offset where each posting list started in the \vbyte\ encoded sequence), and then use \lzend\ to represent it. At construction time, we can obtain different space/time tradeoffs by tuning the {\em delta-codes-sampling} parameter ($ds$) of \lzend\ (see \cite{KN:12} for details). In our experiments we set it to $ds=\{4,16,64,256\}$. At intersection time, we first recover the \vbyte\ encoded sequences and then proceed {\em merge}-wise as in \vbyteLZMA. 

	\item  \repairNo-based representations. As in the \lzend-based representation we compress all the posting lists as a whole, yet we directly work on the sequences of {\em gaps} rather than on their \vbyte\ encodings. We use the grammar-based compressor \repair\ that recursively replaces the most frequent pair of symbols (initially called {\em terminals}, which are those {\em gaps} from the concatenation of all the posting lists) by a new {\em non-terminal} symbol not occurring before in the original sequence. Re-pairing process ends when no pair occurs at least twice. The result is a sequence that could contain both terminal and non-terminal symbols, and a dictionary of substitution rules which is represented in a compact format. This makes up our basic \repairNo\ representation. It allows us to extract individual posting lists,\footnote{To avoid creating Re-Pair phrases that span along two different posting lists, we add a non-repeating separator to mark the beginning of each posting list.} and to perform intersections in a {\em merge}-type fashion over the compressed representation of the lists (non-terminals are expanded/decoded recursively during the traversal of the lists). In addition, in our implementation we added a parameter $repairBreak$ that allows us to stop the recursive pairing when the gain in compression ratio (reduction of size of the compressed sequence) of the current step with respect to the previous step is smaller than $repairBreak$.\footnote{Being $n$ and $m$ respectively the length of the original sequence and the length of the compressed sequence, we initially set $prevRatio \leftarrow (100.0*m/n)$. Then, after each  substitution,  if  $((prevRatio - (100.0*m/n)) < repairBreak)$ we break the Re-pairing process. Otherwise, we simply update $prevRatio \leftarrow (100.0*m/n)$.} In practice, for \repairNo\ we set $repairBreak=0.0$ so that the Re-pairing process is not broken at all.
	
	The compact representation of the dictionary of rules allows us to add additional {\em skipping} data (i.e. the sum of all the {\em gaps} included below that non-terminal symbol) with little space cost. This skipping data allows us to improve both decompression time for a list, and  intersection time. We adapted the {\em merge}-based algorithm from \repairNo\ so that it is boosted by using this skipping data. The resulting algorithm was named {\em skip}. This makes a new representation of posting lists referred to as \repairSkip. In our experiments, we tuned \repairSkip\ with $repairBreak=4\times 10^{-7}$ and  $repairBreak=5\times 10^{-7}$ respectively for the non-positional and positional scenarios.
	
	Initially, neither \repairNo\ nor \repairSkip\ had sampling to speed up intersections, and only \repairSkip\ could benefit of the additional data to boost the intersections. However, since \repairSkip\ was shown to outperform \repairNo, we also created two \repairSkip\ variants using both \CM-type and \ST-type sampling. They are named \repairSkipCM\ (where we set $k=\{1,64\}$) and \repairSkipST\ (with $B=\{ 256, 1024\}$).

	\item \texttt{Ottaviano\&Venturini's Partitioned Elias-Fano indexes}. In \cite{OV14}, Ottaviano and Venturini presented several posting list representation alternatives based on Elias-Fano codes. We used the best of them (\efopt). The source code is available at authors' website.\footnote{\url{https://github.com/ot/partitioned_elias_fano.}} In addition, they included in their framework implementations of some well-known representations such as \optpfd\ (optimized \pfordelta\ variant \cite{YDS09}), \interpolative\ (Binary Interpolative Coding \cite{MS00}), and \varint\ (SIMD-optimized Variable Byte code \cite{Stepanov:2011}). All of them were compared in our parent paper.

\end{itemize}

\paragraph{Our implementation of non-positional and positional inverted indexes} 

Given a text $T$ that is composed of a set of documents, to create our inverted indexes, we process $T$ to gather the different words/terms in $T$, and for each term we obtain the positions were they occur.
The text itself can  be processed with a suitable compressor in order to reduce space needs. In our implementation, we used \repair\footnote{We have also the option to use: a \lzend\ compressor; keeping the  text in plain form; or even not storing the text at all.} as text compressor, which reached compression ratios below $2$\% (we kept the rules uncompressed -as a pair of integers- to speed up decompression), and we added sampling for every \{$1, 2, 8, 32, 64, 256, 4096$\} symbols of the final \repair\ sequence to allow decompressing arbitrary parts of the original text.\footnote{This was of interest in the positional scenario where, as we will see in the next section, we are interested in comparing the snippet extraction time with that of self-indexes.} Yet, the main focus of our parent paper was set on how to efficiently deal with the set of posting lists. 
We included all the above posting lists representations, with the exception of Ottaviano\&Venturini's implementations, within a package that provides us methods to build a compressed representation for a set of posting lists, extract a list, intersect $2$-$n$ lists, show the compressed size, etc.  We did this to obtain compressed representations for posting lists to be used in a non-positional scenario, where we indexed document-ids and intersections were used to solve AND-conjunctive queries for the elements of our query patterns, but also for a positional scenario, where those patterns were taken as phrase queries. In this positional scenario, we indexed actual word/term positions and solving phrase queries implied that the intersections had to consider the order of the terms in the query patterns to return the documents where they occur at consecutive positions.
The building process is handled by a \textbf{build} program that processes the source text and saves all the required structures to disk. A \textbf{searcher} program is on charge of loading those structures into main memory and then allowing us to perform queries. Both programs are linked a posting list representation and a text compressor.

According to the experiments reported in our parent paper, our compressed inverted index variants allow us to perform two types of queries:
\begin{itemize}
	\item {\tt locate(p)} returns the positions of all the occurrences of $p$ in $T$. In the non-positional scenario our experiments (see Section \ref{experiments}) were focused on comparing both the fetch time and the intersection time of the different posting lists representations, so basically  {\tt locate(p)} returns the documents where $p$ occurs. 
	
	In the positional scenario, {\tt locate(p)} returns a pair of the form $(doc, pos\_in\_doc)$. Note that our posting list representation indexes {\em absolute} positions in $T$ rather than pairs $(doc, pos\_in\_doc)$. Therefore, to support returning relative positions within each document, we enhanced our inverted index with an array of offsets that mark the beginning of each document in $T$, and provided an operation {\tt merge-occs-to-docs} which efficiently maps a sequence of absolute positions into pairs $(doc, pos\_in\_doc)$. Therefore, to solve {\tt locate(p)} we initially perform an intersection to gather the positions in $T$ where $p$ occurs, and then perform {\tt merge-occs-to-docs} operation to obtain the final output.
	\item {\tt extract(a,b)} retrieves the text fragment in $T[a,b]$. This is efficiently supported by using the regular sampling added to the (\repair) text representation.
\end{itemize}


\subsubsection{Self-Indexes}\label{sec:self}

A {\em self-index} \cite{Nav:16} is a compact data structure that enables efficient searches over an 
string collection (called the text, $T$), and also replaces the text by supporting extraction of 
arbitrary snippets. Thus, self-indexes provide, at least, two basic queries:

\begin{itemize}
  \item {\tt locate(p)} returns the positions of all the occurrences of $p$ in $T$.
  \item {\tt extract(a,b)} retrieves the text fragment in $T[a,b]$.
\end{itemize}

This functionality allows self-indexes to be considered as an alternative choice to positional inverted
indexes. 

Our original setup comprised three families of self-indexes that were able to capture high repetitiveness. All of 
them have a representative proposal which regards $T$ as a sequence of characters, but two word-oriented
approaches (which process $T$ as a sequence of words) were also proposed. It is worth noting that, in all 
cases, {\tt locate(p)} returns {\em absolute} positions in $T$ that must be then converted into 
(document,offset) pairs. The aforementioned {\tt merge-occs-to-docs} position-document mapping is used again
for such purpose.

\paragraph{CSA-based Self-Indexes}
The Compressed Suffix Array (CSA) by Sadakane \cite{Sad03} is one of the pioneering self-indexes and
proposes a succinct suffix array (SA) encoding. It retains the original locate functionality of the
suffix array (based on binary search), while providing snippet extraction in compressed space. CSA
encompasses two main structures: a bitmap $B$, which marks where the first symbol of the suffixes 
changes in SA, and $\Psi$, an array which stores the position in SA pointing to the next character of a suffix.
$\Psi$ is highly compressible, but it is also massively used to decode a portion of the text. 
Thus, self-indexes dealing with highly repetitive collections must balance $\Psi$ compression and
decoding efficiency to be competitive with respect to compressed positional inverted indexes.

Two different CSA-based approaches were evaluated in our benchmark: \rlcsa\ and \wcsa.

\begin{itemize}
	\item \rlcsa. The Run-Length Compressed Suffix Array, by M\"akinen et al. \cite{MNSV:10}, exploits that 
	   $\Psi$ contains long runs of successive values in highly repetitive collections. It performs 
	   run-length encoding
	   of $\Psi(i)-\Psi(i-1)$ and stores regular samples to allow efficient access to absolute $\Psi$
	   values. This {\em sample} value must be provided to build \rlcsa, and yields different space/time 
	   tradeoffs: larger {\em sample} values are used to reduce space requirements, but it penalizes
	   data access speed; on the contrary, small {\em sample} values increase efficiency at the price
	   of less compression. The authors proposed {\em sample}$=512$, but our experiments consider 
	   7 different values of the form $2^i$, from $i=5$ ({\em sample} $=32$) to $i=11$ ({\em sample} $=2048$), 
	   to analyze particular tradeoffs. A {\em blocksize} parameter is also required to configure internal 
	   bit vectors blocks (it sets the number of reserved bytes per block). We use {\em blocksize} $=32$, as 
	   suggested  by the authors.
	\item \wcsa. The Word Compressed Suffix Array, by Fari\~na et al. \cite{FBNCPR12} adapts CSA to cope with particular features
	   of natural language; i.e. it processes the input text as a sequence of words instead of characters.
	   \wcsa\ transforms the original text into an integer sequence where each position refers to a word/separator,
	   but it does not provide any particular optimization to manage highly-repetitive texts. We included it
	   in our original experiments because it acts as a bridge between inverted indexes and more sophisticated 
	   self-indexes, both in space and time complexity.
	   \wcsa\ builds $B$ and $\Psi$, but over the integer sequence, and provides word-based location and
	   extraction. This requires keeping samples of $SA$ and samples of the inverse of $SA$ (indicating which
	   position of $SA$ points to the $j$-th word) at regular intervals. The non-sampled values can be
	   recovered with $\Psi$. Therefore, three different parameters must be provided at construction time: $\langle sPsi, sA, sAinv\rangle$.	   
	   Note that the inverse of $SA$ is only needed for extract, $SA$ is used both for extract and locate, and $\Psi$ is used in all search operations. Therefore, even though we can tune different setups yielding similar space requirements, it is worth to keep a rather small value of $sPsi$, a larger value of $sA$, and an even larger value of $sAinv$. We evaluated seven different configurations for the sampling 
	   parameters ranging from $\langle sA, sAinv,sPsi \rangle = \langle 8,8,8 \rangle$ to $\langle 2048,2048,2048 \rangle$. In particular, we
	   used the values: $\{ \langle 8,8,8 \rangle, \langle 16,64,16 \rangle, \langle 32,64,32 \rangle, \langle 64,128,32 \rangle,  \langle 128,256,128 \rangle, \langle 512,512,512 \rangle,  \langle 2048,2048,2048 \rangle   \}$.
\end{itemize}

\paragraph{SLP-based Self-Indexes}
We previously motivated that grammar-based compression is a good choice for posting list encoding, but it is also
promissory for self-indexes. Our original paper focused on a particular type of grammar compressor built around
the notion of straight-line program (SLP).\footnote{SLPs are grammars in which a rule $X_i$ generates i) a terminal
$j$, or ii) a pair of non terminals $X_lX_r$.} We explored two SLP-based self-indexes, referred to as \slp and
\wslp, which were carefully tuned for our experiments.

\begin{itemize}
	\item \slp. Claude et al. \cite{CFMPN10} proposed originally a character oriented SLP self-index to
		manage (highly repetitive) biological databases, and then optimized it to cope with natural language collections
	    \cite{CFMPNcikm11}. \slp\ indexes the set of rules as a labeled binary relation, and the reduced
	    sequence (obtained by RePair) using a varied configuration of bit-based structures. \slp\ requires
	    space proportional to that of a Re-Pair compression of the text, but its performance is less competitive
	    than that reported by other self-indexes. The original \slp\ implementation was improved in our parent paper,
	    where we tunned some of its algorithms and internal data structures. A single $q$ parameter is required to
	    build an SLP index; it sets the lengths of the q-grams that are indexed, by an internal dictionary,
	    to improve prefix/suffix searches. $q$ is set by default to $4$ characters because no relevant
	    improvements have been found for larger values.
	\item \wslp. The word-oriented SLP was proposed in our parent paper, and adapted  \slp\ to 
	    perform on a sequence of integers (word identifiers). Its internal configuration is similar to that of the 
	    \slp, but it does not use the q-gram dictionary because it was not competitive for words.
	    Thus, \wslp\ exposes a simple build interface which does not require any parameter.
\end{itemize}

\paragraph{LZ-based Self-Indexes}
Finally, we evaluated two different approaches of self-indexing based on two variants of LZ77-like parsing:
\lzindex\ and \lzendindex. In short, an LZ77-like parsing transforms a text into a sequence of phrases,
each one encoding the first occurrence of a substring. Each substring concatenates a maximal substring 
previously used in the text and a trailing character. \lzindex\ and \lzendindex\ uses a similar configuration
of succinct data structures to encode their corresponding structure of phrases, and to allow fast search
and decode capabilities. 

Both self-indexes were originally proposed by Kreft and Navarro \cite{KN:12}, but we tuned them to improve their original 
space/time tradeoffs. Besides, it is worth noting that they support five different 
configurations.\footnote{These configurations are described in our original paper and basically consider different
structures and search algorithms to provide {\tt locate} and {\tt search} functionality.} The {\em default} 
configuration ({\tt Conf.\#1}) is considered if no parameters are provided. It reports the best performance but 
also the worst space. The following parameter configurations are also allowed: {\tt ``bsst ssst''} (for {\tt Conf.\#2}), 
{\tt ``brev''} (for {\tt Conf.\#3}), {\tt ``bsst brev ssst''}  (for {\tt Conf.\#2}), and {\tt ``bsst brev''} 
(for {\tt Conf.\#5}). The latter reports larger space savings at the price of a bit slowdown, but ensuring a 
competitive performance. We chose it in our benchmark.


\subsection{Experimental Results} \label{experiments}

We experimentally studied the space/time tradeoffs obtained when performing \texttt{locate} operation 
with the described posting list representations, in both the non-positional and positional 
scenarios. In the positional scenario we also added a comparison with the proposed  
self-indexes, and we finally included results regarding the time needed to \texttt{extract} snippets.

In this section we provide a summary of the results  that can be obtained using our replication framework \uilib\
(see Section~\ref{sec:docker}).  
In particular, for \texttt{locate(p)} operation we include results for the case where the patterns are
2-word phrases, and for \texttt{extract(range)} we only include results for snippets of around 13,000 characters.
In Section \ref{sec:testdata} we discuss that there are actually  four types of patterns for \texttt{locate} and two
different snippet lengths for \texttt{extract} operation. 
In that section we also discuss details of the text collections used. In addition, we have also included some fixes to errors 
that were detected in the parent paper during the reviewing process.

The time measures included here are referred to CPU user-times and were obtained using our Docker instance in an Intel(R) i7-8700K@3.70GHz CPU  with 64 GB of DDR4@2400MHz memory. 

\subsubsection{Fixes to the Parent Paper}
Before delving into the details of the experiments described in the parent paper, it is worth noting that developing \uilib\ helped us to
detect some minor errors in the results reported in the parent paper. More precisely:

\begin{itemize}
	\item \qmx\ performance was wrong in the experiment dealing with {\tt locate} operation for 5-words phrase patterns (Figure 3 of the parent paper) in the non-positional scenario. 
		  Its actual performance is slightly faster than the one reported by \vbyteB, but it means that it is half an order of magnitude slower than the fastest choice: \vbyteSTB.
		  The {\tt locate} times of \qmx\ were also wrong in the case of word queries in the positional scenario (Figure 6 of the parent paper). Although it remains as the fastest choice in both cases, its difference to \simplen\ or \vbyte\ is drastically reduced.
	\item The \optpfd\ compression ratios are wrong in the experiments for the positional scenario (Figures 6 and 9). Its actual compression ratio is $29.742\%$ (instead of $31.848\%$), so it is more effective than \interpolative\ and \efopt.		  
\end{itemize}

\subsubsection{Non-Positional scenario: Inverted Indexes}\label{experiments:nopos}

For the non-positional scenario we include a comparison of all the compressed inverted index representations discussed in Section~\ref{sec:compII} when considering \texttt{locate} operation. Yet, in this case we only consider fetch/intersection time and skip the parsing time of the query patterns. That is, we assume all the patterns have been mapped into $ids$ before timing starts.

\begin{figure}[htb]
	\begin{center}
		\includegraphics[angle=-90,width=0.49\textwidth]{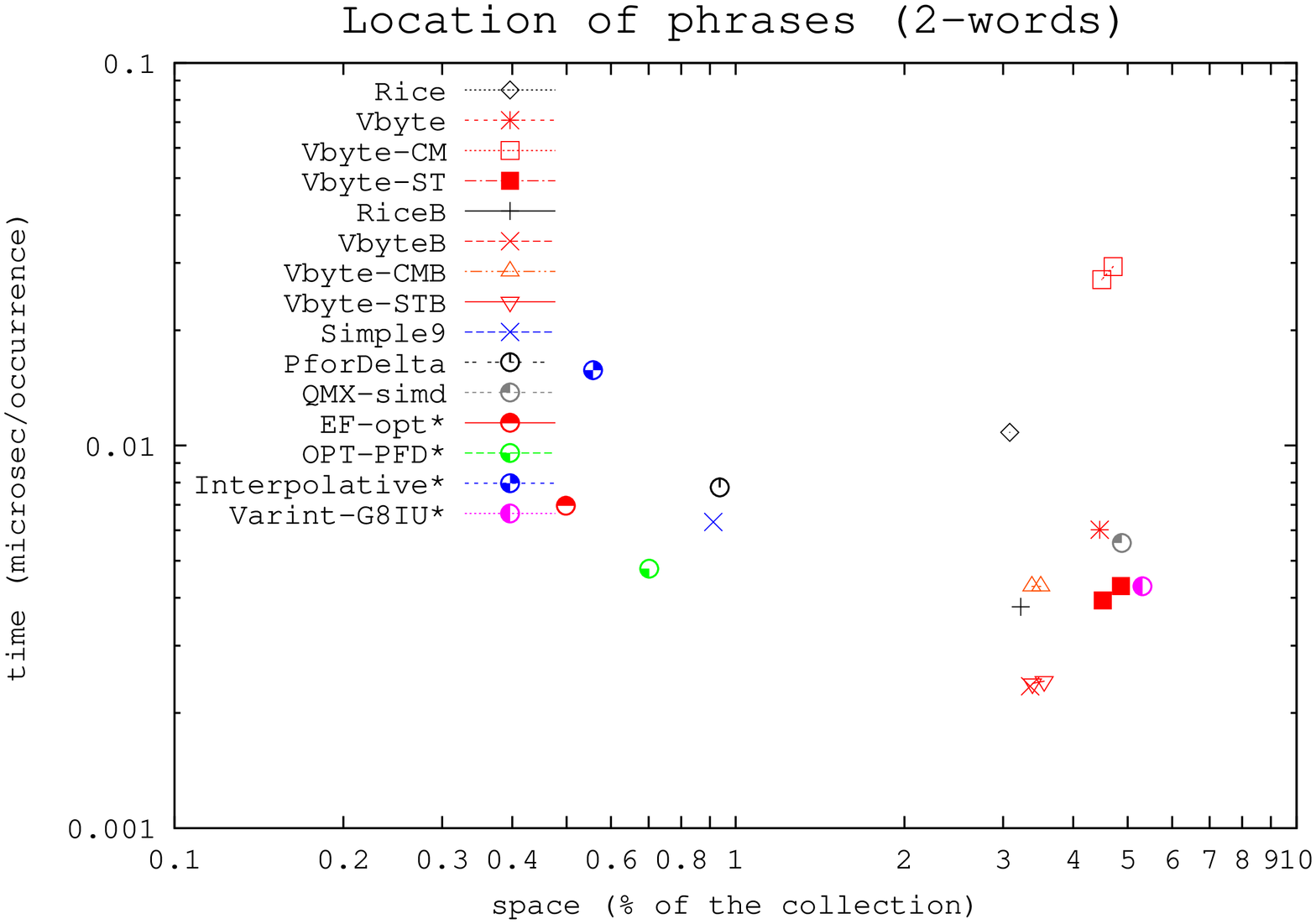}
		\includegraphics[angle=-90,width=0.49\textwidth]{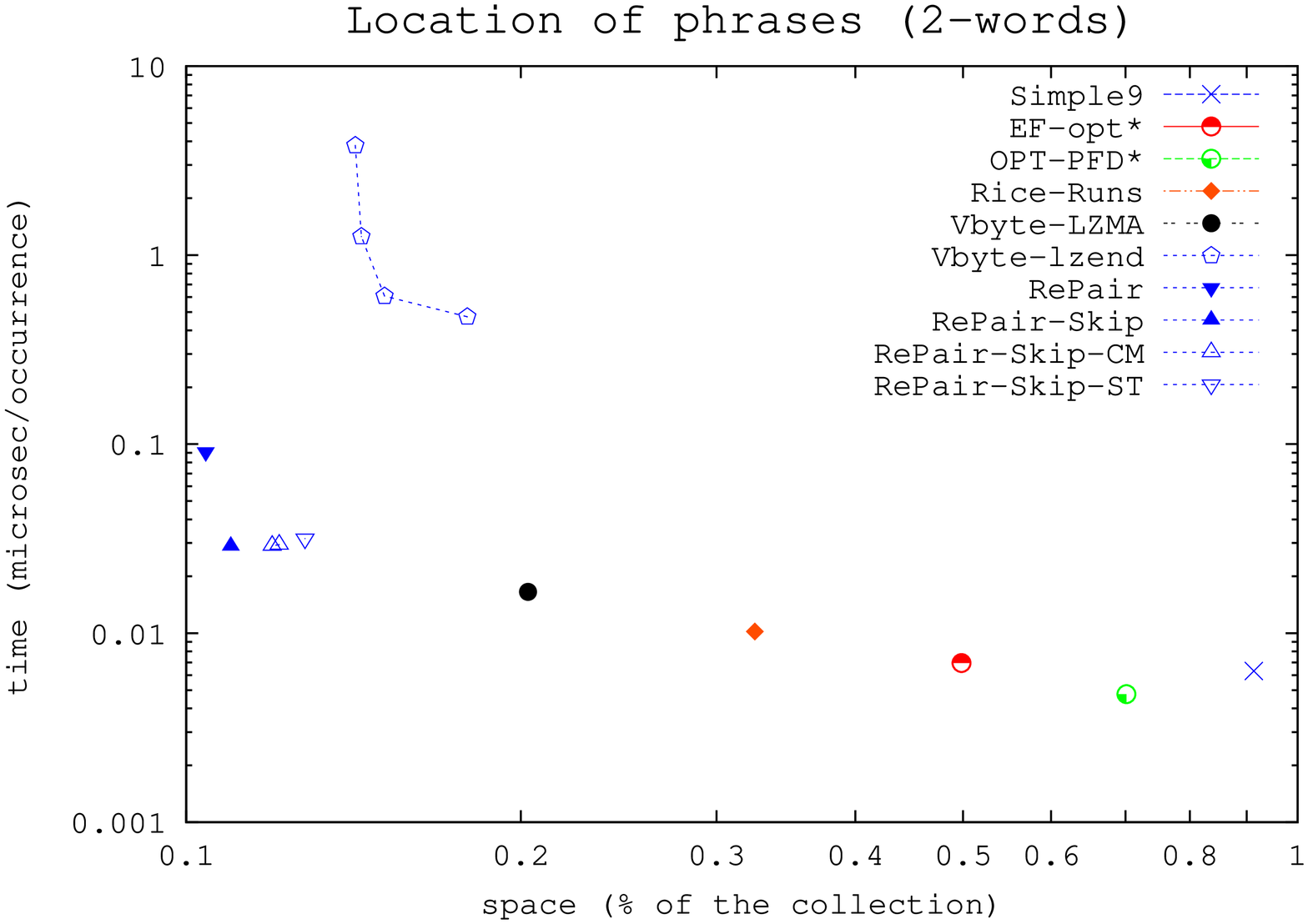}		
		\caption{Main results for the non-positional scenario.}
		\label{fig:exp.nopos}
	\end{center}
\end{figure}

Figure~\ref{fig:exp.nopos} shows the results. In the left part, we show the results for the state-of-the-art techniques. In the right part we also include our proposals. Recall that for the variants from \texttt{Ottaviano\&Venturini's} framework \cite{OV14} we used exactly their source code and simply adapted the format of our data and patterns to run their {\tt build} and {\tt search} programs. Those techniques are marked with an `\texttt{*}' in the figures.


As shown in the parent paper, our new techniques \riceRuns\, \vbyteLZMA, \vbyteLzend, and \repair-based variants drastically reduced the space needs of the existing techniques. Yet, they were typically also much slower (particularly in the case of \vbyteLzend).

\subsubsection{Positional scenario: Inverted Indexes and Self-Indexes}\label{experiments:pos}
In this scenario we compare both our inverted index representations and self-indexes at both \texttt{locate} and snippet \texttt{extract} operations. 

\paragraph{Pattern Location}
We did not include all the inverted index variants used in the non-positional scenario, but only those that could be of interest here.
For the self-indexes, timing includes the time needed to perform \texttt{(locate(p))} operation to obtain the positions where \texttt{(p)} occurs, and then converting absolute positions to $(document,offset)$ pairs using {\tt merge-occs-to-docs} operation. 
For the inverted indexes, timing includes: {\em (a)} the parsing time required to map each pattern into a sequence of $ids$; {\em (b)} performing intersection of the required posting lists (for those $ids$); and {\em (c)} running {\tt merge-occs-to-docs} to obtain the final result. For the techniques from \texttt{Ottaviano\&Venturini's} framework we directly used their sources \textit{with no modifications} to measure {\em (b)} times, whereas time measures for stages {\em (a)} and {\em (c)} where done separately with a program we also implemented. During step {\em (a)}, our program not only measures times, but also outputs the sequence of $ids$ obtained after the parsing of each pattern \texttt{p}. These $id$-based patterns are used to perform intersections with the techniques from \texttt{Ottaviano\&Venturini's} framework.

	
	
\begin{figure}[htb]
	\begin{center}
		\includegraphics[angle=-90,width=0.49\textwidth]{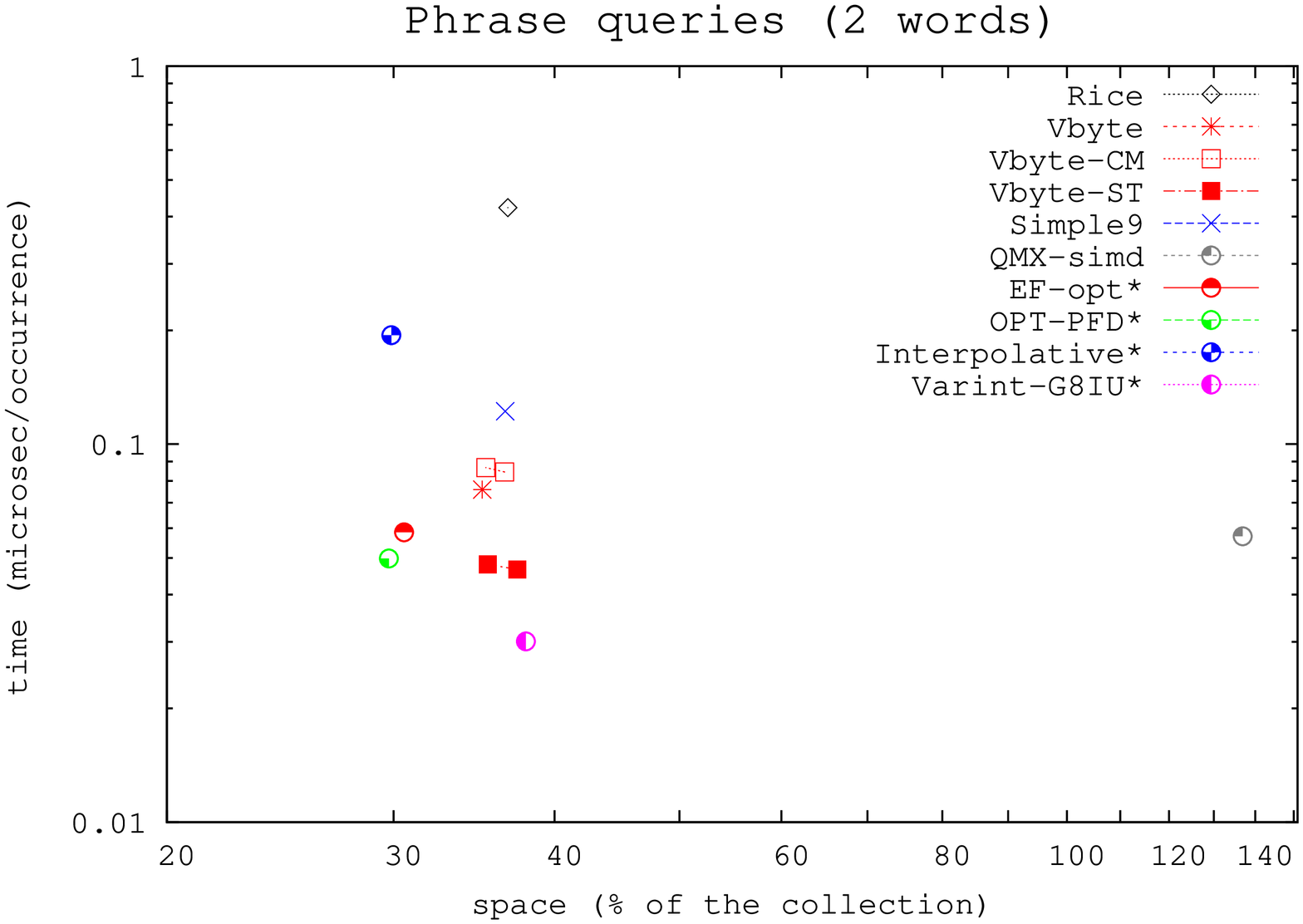}
		\includegraphics[angle=-90,width=0.49\textwidth]{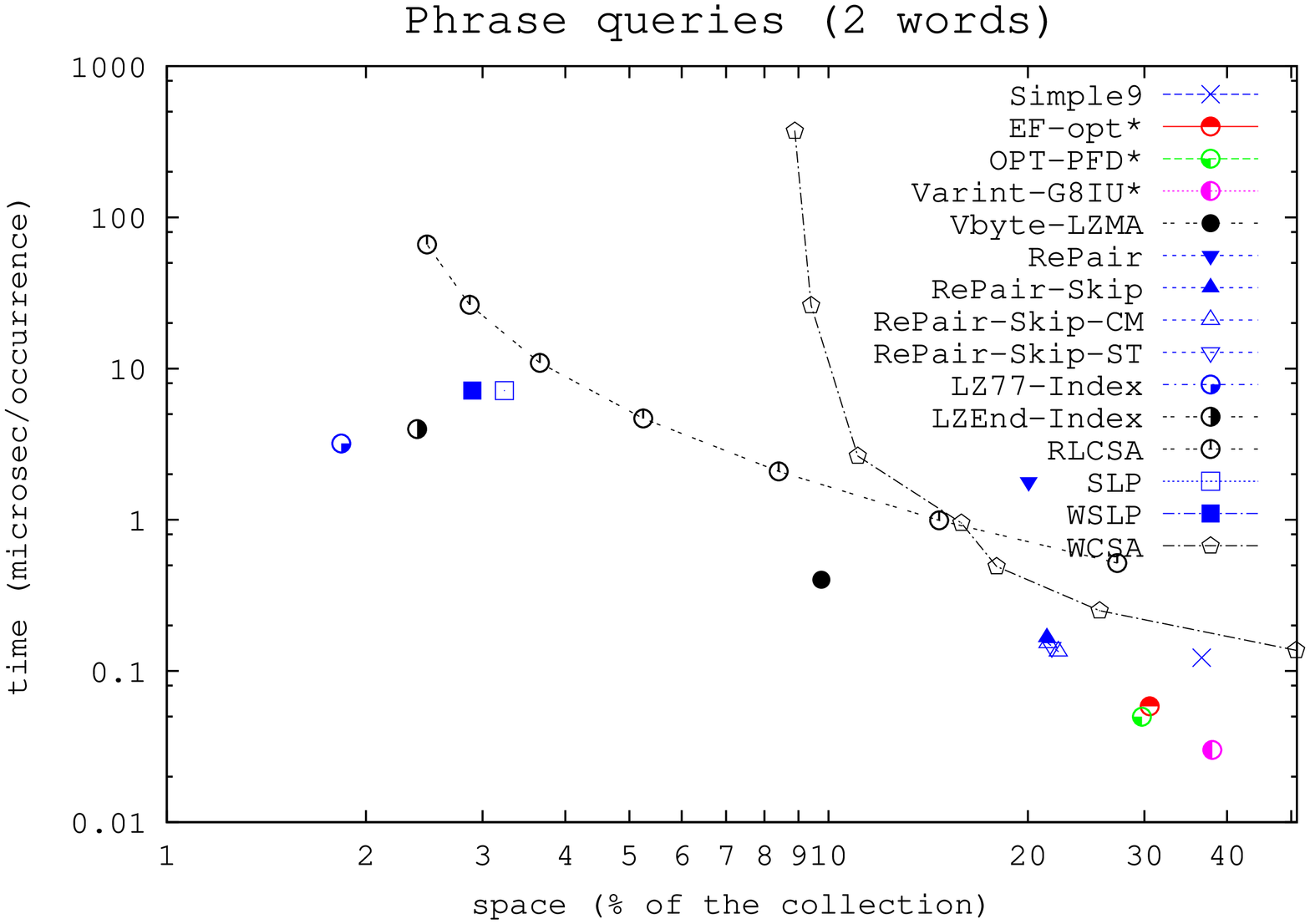}
		\caption{Main results for the positional scenario.}
		\label{fig:exp.pos.loc}
	\end{center}
\end{figure}

Figure~\ref{fig:exp.pos.loc} shows the results. On the one hand, our \vbyteLZMA\ and \repair-based inverted indexes still improve the space needs of the other representations, while being typically  up to one order of magnitude slower. On the other hand, self-indexes showed to obtain around one order of magnitude reduction on space, while becoming up to three orders of magnitude slower than the fastest inverted index alternative.

\begin{figure}[htb]
	\begin{center}
		\includegraphics[angle=-90,width=0.49\textwidth]{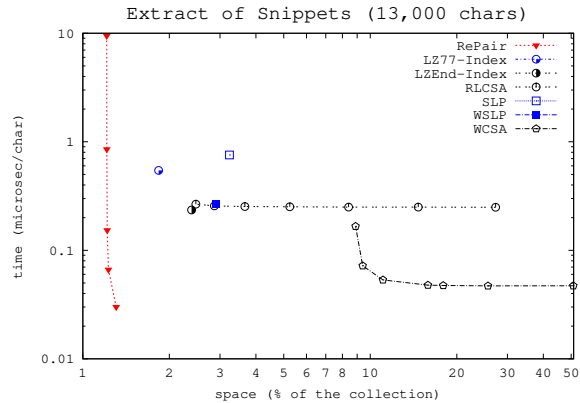}
		
		\caption{Main results regarding text extraction performance for both self-indexes and text compressed with \repair.}
		\label{fig:exp.pos.extract}
	\end{center}
\end{figure}

\paragraph{Snippet Extraction}
We report the time needed to extract text snippets $T[s,e]$ from the self-indexes and compare them with the decoding performance of \repair, that was the text compressor chosen to compress the document collection for the representations using inverted indexes. Recall that \repair\ is coupled with the additional sampling that permit partial decompression, as discussed in Section~\ref{sec:compII}. Results are shown in Figure~\ref{fig:exp.pos.extract}. We can see that \repair\ is very fast at decompression (when a dense sampling is used) and requires less than $1.5$\% of the size of the document collection. With very similar space needs (around $2-3$\%) we find most of the self-indexes (\lzendindex, \lzindex, \rlcsa\, \slp, and \wslp), yet they are typically much slower. Finally, \wcsa\ competes in speed with \repair, but requires around one order of magnitude more space.



\section{The \uilib\ Framework}
\label{sec:3}
Our experimental framework was named  \uilib\ {\em (universal indexes for Highly Repetitive Document Collections)}. It is licensed under the GNU Lesser General Public License v2.1 (GNU LGPL), is hosted at a GitHub repository,\footnote{\url{https://github.com/migumar2/uiHRDC/}} and it is also published through Mendeley Data \cite{FMPCN:18}. It includes all the required elements to reproduce the main experiments in our original paper including datasets, query patterns, source code, and scripts. 

\begin{figure}[ht!]
	\begin{center}
		\includegraphics[angle=0,width=0.65\textwidth]{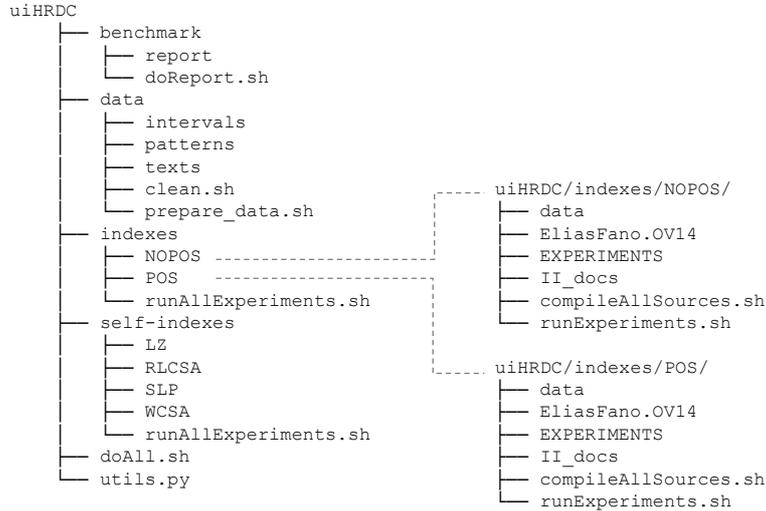}
		\caption{Structure of the \uilib\ repository.}
		\label{fig:github}
	\end{center}
\end{figure}

\begin{figure}[t!]
	\centering
	\includegraphics*[width=.95\textwidth]{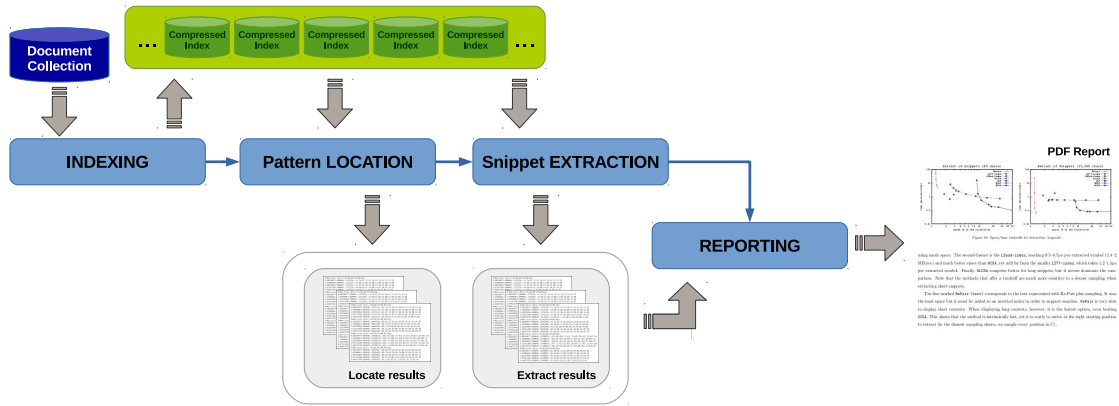} \\ 
	\caption{\label{fig:workflow} Workflow used in \uilib\ to reproduce our experiments.}
\end{figure}

The general structure of the \uilib\ repository is shown in Figure~\ref{fig:github}. It can be seen that, under the root of the repository, we include:
{\em i)} directory \texttt{benchmark} which includes a \LaTeX\ formatted report and a script that will collect all the data files resulting from running all the experiments and will generate a PDF report with all the relevant figures (including those in Section~\ref{experiments}). Further python scripts required for such task are also included in directory \texttt{utils.py}.
{\em ii)} Directory \texttt{data}, which includes the text collections (7z compressed), and the query patterns.
{\em iii)} Directories \texttt{indexes} and \texttt{self-indexes} that contain the source code for each indexing alternative, and scripts that permit to run all the experiments for each technique. This includes the construction of each compressed index of interest (using a \textbf{builder} program) and then performing both \texttt{locate} 
and \texttt{extract} operations over that index (using the corresponding \textbf{searcher} program). Each experiment will output relevant data to a results-data file.
And {\em iv)} a script \texttt{doAll.sh} that will drive all the process of decompressing the source collections; compiling the sources for each index and running the experiments with it; and finally, generating the final report. The overall workflow followed is depicted in Figure~\ref{fig:workflow}.

In the following section we will include further details regarding the contents of our repository.


%


\subsection{Test Data} \label{sec:testdata}
Within \uilib\ we provide in \texttt{data} directory, both the document collections and the query sets used in the experimental setup of our parent paper.
They are described below.

\subsubsection{Document Collections}

Our document collections were created from the 108.5 GB Wikipedia collection described by He et al.~\cite{HZS10}, which contained 10\% of 
the complete English Wikipedia from 2001 to mid 2008. It contains 240,179 articles, and each of them has a number of versions. Its statistics
are shown in Table~\ref{tab:coll}. Note that we did not have the original 108.5 GB collection, but a filtered (tag-free) version of it whose size
totaled 85.58 GB. Therefore, the original collection was $1.27$ times larger than ours.

\begin{table*}[htb]
	{\small
		\begin{center}
			\begin{tabular}{|l|r|r|r|r||r|r|}
				\hline                                                                                           
				Subset   & Size~~ & Articles & Number of & Versions / &       {Filename}         &        {Filesize}  \\
				         & (GB)~  &          & versions  & article~~~ &     (within \uilib)      &        {(GB)}      \\
				\hline                                                                                           
				\hline                                                                                           
				Full     & 108.50 &  240,179 & 8,467,927 &      35.26 &        {\tt --}          &            {85.58} \\
				Non-pos  &  24.77 &    2,203 &   881,802 &     400.27 &        {text20gb.txt}    &            {19.53} \\
				Pos      &   1.94 &    4,327 &   149,761 &      34.61 &       { wiki\_src2gb.txt}&             {1.94} \\
				\hline                                                                                           
				
			\end{tabular}
		\end{center}
	}
	\caption{Detailed statistics of the document collections used.} 
	\label{tab:coll}
\end{table*}

From the {\tt Full} collection of articles, we chose two subsets of the articles, and collected all the versions of the chosen articles. 
For the non-positional setting our subset (\texttt{Non-pos}) contains a prefix of 19.53 GB of the full collection, whereas for positional indexes we
chose 1.94 GB of random articles. However, for a fair comparison with the techniques from~\cite{HZS10}, in the non-positional scenario we scaled the size
of the \texttt{Non-pos} subset using the above $1.27$ factor when referring to its space requirements. Additional statistics of our two subsets are
also included in Table~\ref{tab:coll}.

%

\subsubsection{Query sets}

Since the experiments target at providing space/time for the different indexing alternatives 
when performing \texttt{locate(pattern)} and \texttt{extract(interval)} queries we provide two types
of query sets for each document subcollection.

\begin{itemize} 
	\item Query sets for \texttt{locate}: We provide four query sets, each of them containing 1,000 queries, which include:
	{\em i)} two query sets composed of one-word patterns chosen at random from the vocabulary of the indexed subcollection. In the first
	case ({\em W$_a$}), it includes low-frequency words occurring less than 1,000 times. In the second case, the query set ({\em W$_b$}) includes
	high-frequency words occurring more than 1,000 times; {\em and ii)} two query sets with 1,000 phrases composed of 2 and 5 words that
	were chosen randomly from the text of the subcollection (with no restrictions on its frequency).
	
	\item Interval sets for \texttt{extract}: Aiming at measuring extraction time when recovering snippets of length 80 (around one line) and 13,000 (around one document, in our
	collection) characters, we generated: {\em i)} a set of 10,000 intervals of width 13,000 characters from the {\tt POS} text collection, and {\em ii)}  a set containing
	100,000 intervals of width 80 characters. Since these intervals are not suitable for our word-based self-indexes (\wcsa\ and \wslp), and assuming that the average word length 
	is around $4$ in our datasets, we also generated two additional sets composed respectively of 10,000 intervals containing $3,000$ words each, and 100,000 intervals containing $20$ words each.
\end{itemize}



\subsection{Source Code and scripts}

As indicated above, the source code provided in our \uilib\ repository has two main directories \texttt{indexes} and \texttt{self-indexes}. Those directories include both the source code and the scripts required to reproduce our experiments.
In this section we include more details regarding their structure so that an interested reader can rapidly understand how they are organized.

\subsubsection{Indexes}
\begin{figure}[]
	\begin{center}
		\includegraphics[angle=0,width=1.00\textwidth]{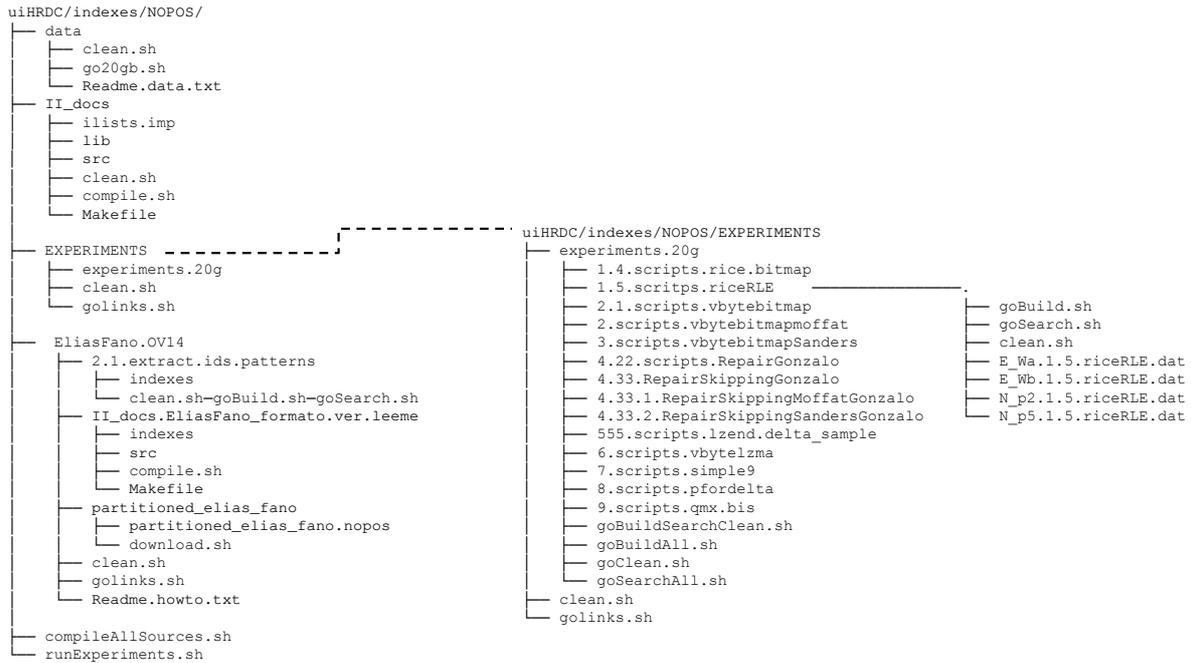}
		\caption{Structure of the \texttt{uiHRDC/indexes} directory in the \uilib\ repository.}
		\label{fig:githubII}
	\end{center}
\end{figure}

Under \texttt{uiHRDC/indexes} directory, as it is shown in Figure~\ref{fig:github}, we can find a script \texttt{runAllExperiments.sh} and two directories \texttt{NOPOS} and \texttt{POS}. Basically, that script enters both directories, and then runs two scripts: one to compile the source codes (\texttt{compileAllSources.sh}) and another one to launch the experiments (\texttt{runExperiments.sh}) in that directory. An interested reader should probably start by opening those small scripts. These scripts are included in Figure~\ref{fig:githubII}, where we show the structure of directory \texttt{NOPOS} (the structure of \texttt{POS} is almost identical).

\paragraph{Directory \texttt{uiHRDC/indexes/NO-POS}: Non-Positional Inverted Indexes}
Under this directory, we can find the scripts \texttt{compileAllSources.sh} and  \texttt{runExperiments.sh} discussed above, a \texttt{data} directory were links to \texttt{uiHRDC/data} will be created by the scripts under this directory, and finally two main parts: {\em i)} \texttt{EliasFano.OV14} with the sources and scripts needed to reproduce our experiments for techniques \efopt, \interpolative, \varint, and \optpfd; and {\em ii)} Directories \texttt{II\_docs} and \texttt{EXPERIMENTS} that include respectively the source code of the remaining inverted index variants and the corresponding scripts to run the experiments. We include more details below.

\begin{itemize}
	\item Our implementation and scripts within directories \texttt{II\_docs} and \texttt{EXPERIMENTS}. We have organized the source code for  our inverted indexes within \texttt{II\_docs} directory. In Figure~\ref{fig:githubII}, we can see a script \texttt{compile.sh} and three subdirectories: \texttt{ilists.imp}, {\tt lib}, and \texttt{src}. The implementation for all our types of compressed posting list representations is contained within directory \texttt{ilists.imp}. Script \texttt{compile.sh} will create a package for each of them and will move such package into \texttt{lib} directory. Finally, the source code for our non-positional compressed inverted index, located within \texttt{src} directory, will be linked with each of those posting list representations to obtain the final \texttt{BUILD*} and \texttt{SEARCH*} binaries for each all our variants of non-positional inverted index. 
	
	In Figure~\ref{fig:githubII}, we can also see the contents of directory \texttt{EXPERIMENTS/experiments20gb}. Basically, there is a subdirectory for each technique with scripts {\tt goBuild.sh} (to create the indexes), {\tt goSeach.sh} (to perform query operations), and {\tt clean.sh} (to clean temporal stuff). Those techniques are respectively (top-to-bottom in the figure): {\em 1.4)} \rice\ and \riceB, {\em 1.5)} \riceRuns, {\em 2.1)} \vbyte\ and \vbyteB, {\em 2)}\vbyteCM\ and \vbyteCMB, {\em 3)} \vbyteST\ and \vbyteSTB, {\em 4.22)} \repairNo, {\em 4.33)} \repairSkip, {\em 4.33.1)} \repairSkipCM, {\em 4.33.2)} \repairSkipST, {\em 555)} \vbyteLzend, {\em 6)} \vbyteLZMA, {\em 7)} \simplen, {\em 8)} \pfordelta, and {\em 9)} \qmx. In addition, a script {\tt goBuildSearchClean.sh} is in charge of entering each directory running the experiments for the corresponding technique. The output of those experiments are data files containing both space and time statistics (see files {\tt E\_Wa.1.5.riceRLE.dat, ...}).

	\item Ottaviano\&Venturini's variants within directory \texttt{EliasFano.OV14}: This directory contains: {\em i)} A subdirectory {\tt partitioned\_elias\_fano} containing the source code from Ottaviano\&Venturini's framework and a script to run the experiments corresponding to variants \efopt, \interpolative, \varint, and \optpfd; and {\em ii)} two additional directories {\tt II\_docs.EliasFano\_formato.ver.leeme} and {}{\tt 2.1.extract.ids.patterns} that contain source code and scripts to transform the text-based patterns into the $id$-based patterns that will be used at query time. In this case, the output of the experiments is written to a log-file and finally a Phyton script parses such file to gather the values regarding space and time measures for each technique.
	
\end{itemize}

\paragraph{Directory \texttt{uiHRDC/indexes/POS}: Positional Inverted Indexes}

As indicated above, the structure of this directory is almost identical to that of directory {\texttt{uiHRDC/indexes/NO-POS}}. Therefore, we include no further details here.

\subsubsection{Self-Indexes}

\begin{figure}[]
	\begin{center}
		\includegraphics[angle=0,width=0.60\textwidth]{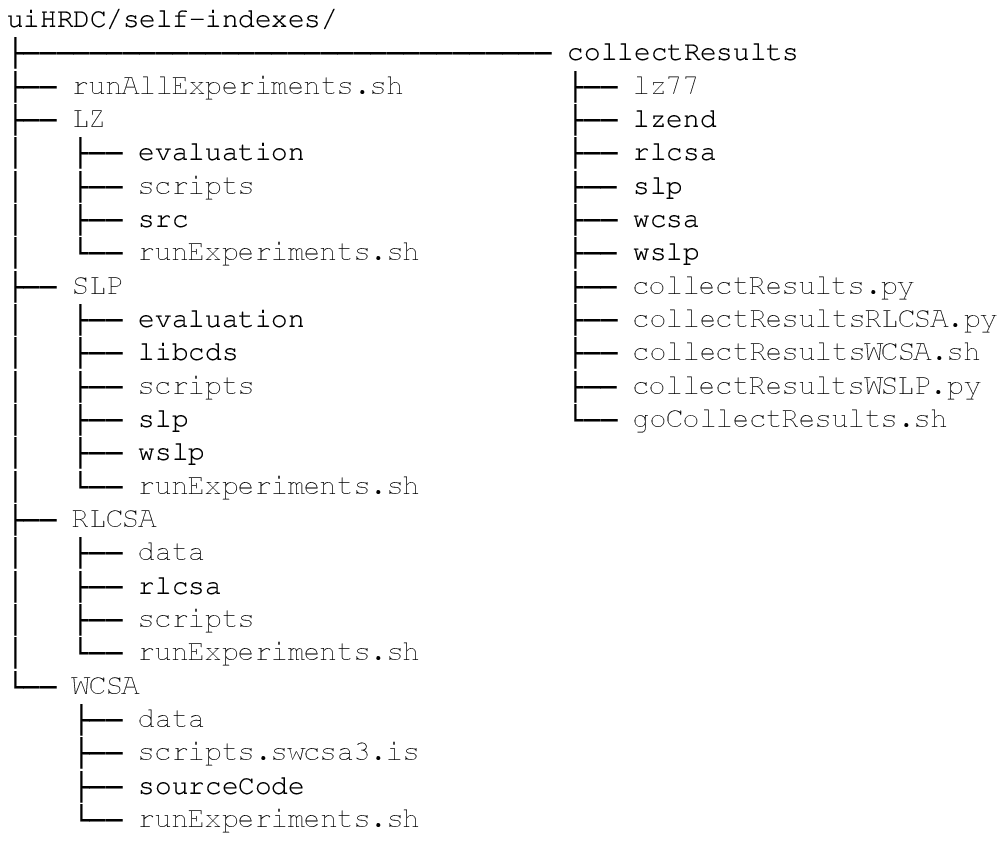}
		\caption{Structure of the \texttt{uiHRDC/self-indexes} directory in the \uilib\ repository.}
		\label{fig:githubSelfindexes}
	\end{center}
\end{figure}

Under directory \texttt{uiHRDC/self-indexes}, as it is shown in Figure~\ref{fig:githubSelfindexes}, we can find a script \texttt{runAllExperiments.sh} and one directory for the different types of self-indexes. In particular, we find directories {\tt RLCSA}, {\tt WCSA},   {\em LZ } (for \lzindex\ and \lzendindex) and {\tt SLP} (for \slp\ and \wslp). In those directories we will find, among others, the source code and scripts to run the experiments for each technique. 

In addition, directory \texttt{uiHRDC/self-indexes/collectResults} contains scripts to process the log files obtained when we run the script {\tt runAllExperiments.sh} and create gnuplot-type data files for each technique that are used to make up the final report for the experiments.

%
%
%
%
%

\section{Deploying the Experimental Setup with Docker}  
\label{sec:docker}

For those readers interested in reproducing our experiments in their machine, we provide a {\em Docker} environment that 
will allow them to:
\begin{enumerate}
	\item Reproduce our test framework. We create a docker image with Ubuntu 14 (ubuntu:trusty) that includes all the libraries  and software requirements to compile/run our indexing alternatives and also build the final report. These includes packages that are installed via \texttt{apt} such as:
	gcc-multilib, g++-multilib, cmake, libboost-all-dev, p7zip-full, openssh-server, screen, texlive-latex-base, and texlive-fonts-recommended; and finally, snappy-1.1.1\footnote{\url{https://github.com/google/snappy}}, which we included in a snappy-1.1.1.tar.gz file, and gnuplot-qt, which in included in gnuplot-4.4.3.tar file.
	In addition, the contents of the \uilib\ framework (downloaded from our repository) and copied into \texttt{/home/user/uiHRDC} directory of our docker image.

	\item Connect to an instance of our docker image using \texttt{ssh/sftp}. Basically, we \textbf{expose} port 22 and create a user \texttt{'user'} with password \texttt{'userR1'} who can connect via \texttt{ssh}. This will allow the reader to connect to the docker container as to any remote server and to retrieve the final report by sftp. Once connected,  \texttt{user} has \texttt{sudo} priviledges and, for example, can become {\tt root} simply entering {\em sudo su}. 
	
	\item Run \texttt{doAll.sh} script to automatically run all our experiments and generate our final report. This script must be run by \texttt{root} user. Note that we have installed (via apt-get) \texttt{screen}  virtual terminal so that the user can disconnect from the docker container and still keep {\tt doAll.sh} script running. 
\end{enumerate}

The minimum hardware requirements to run \texttt{doAll.sh} script would be to have a machine with at least 32GB RAM (and 16GB swap) and around 200GB of free disk space (in the file system were Docker keeps its files\footnote{If you experiment space issues, a simple workaround could be to check which is the directory where Docker creates its images (e.g. {\tt /home/shared/docker/tmp/} or {\tt /var/lib/docker}), and to replace it by a symbolic link pointing to a location in a larger partition/disk.}). In our machine, i7-8700K@3.70GHz CPU (6 cores/12 siblings) with 64 GB of DDR4@2400MHz memory and a 7200rpm SATA disk, it took around 40 hours to run all the experiments from \texttt{doAll.sh} script.

\subsection{Running the experiments: step-by-step} 
Below we show all the commands needed to use Docker\footnote{To install Docker, please refer to the installation guide for your host operating system: {\tt https://docs.docker.com/install/}. In our Ubuntu system it simply consisted in running ``{\tt sudo apt-get install docker.io}''} to reproduce our framework within a Docker container, then run the experiments within that container, and retrieve the final report (including also the gnuplot-type result data files).
\begin{enumerate}
	\item Create a temporal working directory (i.e. \texttt{\$TMP}) and move to it: \texttt{mkdir \$TMP} and \texttt{cd \$TMP}

	\item Clone the \uilib\ GitHub repository at \url{https://github.com/migumar2/uiHRDC/}. It will create a directory \texttt{\$uiHRDC}. Please rename it as \texttt{\$DIR}. Then move to directory \texttt{\$DIR}. Within \texttt{\$DIR} you will find a file \texttt{Dockerfile} and two directories: \texttt{docker} and \texttt{uiHRDC}. 
	
		\texttt{\$TMP> git clone https://github.com/migumar2/uiHRDC.git} \\
		\texttt{\$TMP> mv uiHRDC \$DIR } \\
		\texttt{\$TMP> cd \$DIR}  
	
	
	\item Creating a docker image named {\tt repet}: (you must be at \texttt{\$DIR} directory) \\
		\texttt{\$DIR> sudo docker build -t repet --rm=true . }~~~~ \# Note: there is a final `dot' (.)  
		
	\item Running a docker container named \texttt{repet-exp} that exports ssh/sftp port (22) via port \texttt{22222}: \\
		\texttt{\$DIR> sudo docker run --name repet-exp -p 22222:22 -it repet} \\
		Now, you can press \texttt{CTRL+P CTRL+Q} to detach from docker-container.
		
	\item Connecting via ssh to the container and becoming {\tt root}.
	
		\texttt{\$DIR> ssh user@localhost -p 22222}  ~~~~~\# enter password {\tt userR1} when prompted.\\
		\texttt{\$user@docker> sudo su } ~~~~~~~~~\# enter password {\tt userR1} when prompted.\\
		\texttt{\$root@docker> screen -t EXPERIMENTS } ~~~~\# optional, to launch screen virtual terminal.
		
		
	\item Move to the working directory {\tt /home/user/uiHRDC} where {\tt doAll.sh} script is located.\\
		\texttt{\$root@docker> cd /home/user/uiHRDC }
		
	\item Now we run {\tt doAll.sh} script.
		\texttt{\$root@docker> sh doAll.sh } \\
		and wait until {\tt doAll.sh} had completed. 

	\item If you want to retrieve the final report and the results (gnuplot-type data files) via sftp you must grant access to those files to user {\tt user}.\\
		\texttt{\$root@docker>tar czvf /home/user/uiHRDC/report.tar.gz /home/user/uiHRDC/benchmark } \\
		\texttt{\$root@docker>chown user:user /home/user -R }  ~~~~~

    \item Now we can close the ssh session (or detach from the docker container).\\
		\texttt{\$root@docker> exit }  ~~~~~\\
		\texttt{\$user@docker> exit }  

    \item Now we can connect by  sftp or scp using again port 22222 to download {\tt /home/user/uiHRDC/report.tar.gz} from the
    docker container.

		\texttt{\$DIR> sftp -P 22222 user@localhost:/home/user/uiHRDC}  ~~\# enter password {\tt userR1} when prompted.\\
		\texttt{\$sftp> get report.tar.gz } 
		
		or alternatively: \\
		\texttt{\$DIR> scp -P 22222 user@localhost:/home/user/uiHRDC/report.tar.gz .}~~\# use password {\tt userR1}
		
		When decompressed, you will find the report here {\tt \$DIR/benchmark/report/report.pdf } and all the gnuplot data files within  {\tt \$DIR/benchmark/report/figures} directory.

	\item Finally you can stop {\tt repet-exp} container, and if needed remove it and also {\tt repet} image:\footnote{The reader can use \texttt{sudo docker ps} and \texttt{sudo docker images} to see respectively the active containers and existing images.} \\
		\texttt{\$DIR> sudo docker stop repet-exp } \\
		\texttt{\$DIR> sudo docker rm repet-exp } \\
		\texttt{\$DIR> sudo docker rmi repet } 
				
\end{enumerate}


\section{Conclusions and Future Work}
\label{s:conc}
We have briefly described all the techniques used in our original paper. In total, we had twenty two variants
of posting list representations available that were used to create both non-positional and positional inverted indexes. In addition,
we had six self-indexing techniques. Since those techniques have their own configuration parameters, and in some cases dependencies of libraries/software, it would be hard to replicate the results and to reuse our techniques by simply reading our original paper and cloning the (GitHub) repository were we had made our sources available. To overcome those limitations, this paper includes a detailed description of our replication framework \uilib. An interested reader could find not only our source code, our document collections and the query patterns used in our experiments, but also a set of related scripts. Those scripts permit us to replicate all our experiments with little effort and, additionally, generate a PDF report (using python, gnuplot, and latex) that contains the figures with the experimental results from our parent paper. In addition, we provide some configuration files to create a Docker container that reproduces our test environment (including all the required dependencies), and instructions regarding how to start the container, run the experiments, and finally download a copy of the final report from the Docker container. We hope that the descriptions and instructions provided  along this paper will simplify the work of any reader interested in reusing the experiments from our original paper.

An interesting line of future work is to redesign \uilib\ to facilitate that new indexes to be added to the framework. The resulting benchmark framework would ensure reproducible experimentation for ongoing research about compression and indexing of highly repetitive collections.


\section{Revision Comments}
\label{s:revision}
We would like to thank the authors for providing this valuable reproducibility platform, which allows both reproducing the results of its parent paper and encouraging the evaluation of new indexes for highly repetitive document collections in an easy and systematic way. Undoubtedly, this will be a topic of active research in the coming years given, for example, the cheapening of gene sequencing technology. Hence, the public availability of this type of benchmarking platform is becoming imperative. For this reason, we sincerely expect that uiHRDC sets an experimental standard for this line of research.

On the other hand, this work confirms again that the production of reproducible science is a difficult task, thus reproducibility of any research work must be considered and planned since the very beginning stages of development. Despite the reviewers reached a consensus about the reproducibility of the paper, the rigorous review process raised some reproducibility issues that left us some significant lessons on the difficulties of producing fully reproducible science and developing such ambitious reproducible platform as introduced herein. Next, we discuss our experience with this work as well as the lessons learned from our review.

We fully support the choice of using Docker for this kind of reproducibility experiments. Using a container makes it extremely simple to reproduce the experimentation framework, which otherwise would require to install the dependencies and compile the tested codes manually. However, review process unveiled other issues on the set-up and running of the uiHRDC experiments, which were fixed by providing specific instructions solving them. For instance, authors added information in the output report to warn the experimenter in those cases in which the experiments failed by lack of resources, as well as to ensure that the folder used by Docker resides in a drive with enough memory, which might not be the case for the default drive used by Docker.   

In addition to the aforementioned set-up and running issues, some others arose with the experimentation itself in a first review: (1) a significant mismatch in the ordering of time-axis results derived from the uncertainty in the evaluation of running time values; (2) a difference in the compression ratios obtained for the OPT-PFD indexing method; (3) missing values for Vbyte-Lzend and QMX-SIMD methods derived from software problems which did not provide any warning to the experimenter; and (4) differences of scaling between the original figures and those generated by uiHRDC, which made a detailed checking process difficult. Subsequently, all experimentation issues detailed above were either fixed or honestly and rigorously justified in the paper in a sample of best research practices. First, time measurement was significantly improved by increasing the number of evaluations with the aim of reducing its fluctuation. It worths to highlight that special attention should be put in any research for the reproducibility of running time values and their conclusions. Second, a few code bugs were fixed and the output experimentation report was extended to provide detailed information on the execution of the experiments. This further information shown to be very useful to guide the review process, and also to detect possible bugs for particular execution environments. Finally, axis scaling was revised to match the original paper exactly, and a bug with gnuplot was fixed so the generated report is produced with the same symbology of the parent paper. These later two issues highlight the importance of the presentation of the results to properly communicate research findings in a clear way.

The uiHRDC framework is a first try of standardizing experiments that, in absence of such a framework, would need to be repeated each time a new index is introduced in the literature. Thus, uiHRDC will be a very valuable resource to the research community by avoiding this tedious and costly task, which is also prone to errors. 

As forthcoming activities, we invite the authors to extend uiHRDC with the aim of removing some drawbacks that hinder the integration of new indexes in their platform. The main drawback is that adding a new index is not trivial and requires editing several scripts; thus, any improvement in this sense will contribute to turn uiHRDC into a standard for comparison by future researchers.

\section*{Acknowledgments}
This paper is funded in part by European Union's Horizon 2020 research and innovation programme under the Marie Sklodowska-Curie grant agreement No 690941 (project BIRDS). 
Antonio Fari\~na is funded by Xunta de Galicia/FEDER-UE [CSI: ED431G/01 and GRC: ED431C 2017/58]; by MINECO-AEI/FEDER-UE [ETOME-RDFD3: TIN2015-69951-R]; and by MINECO-CDTI/FEDER-UE [INNTERCONECTA: uForest ITC-20161074].
Miguel A. Mart\'inez-Prieto is funded by  MINECO-AEI/FEDER-UE [Datos 4.0: TIN2016-78011-C4-1-R].
Gonzalo Navarro is funded by the Millennium Institute for Foundational Research on Data (IMFD), and by Fondecyt Grant 1-170048, Conicyt, Chile.

\bibliographystyle{abbrv}
\bibliography{repro}

\appendix
%
%

\section{Compression Ratios} \label{app:A}
This appendix shows the exact compression ratios for all the techniques included in our \uilib\ framework. The corresponding tuning parameters are provided for each case ($\times$ is used for those techniques that do not require any parameter). Those values are included in Table~\ref{comp:indexes}. 

Note that in the positional scenario, when the source text collection is compressed with \repair, the parent paper allowed the extraction of snippets consisting of either $80$ or $13,000$ chars. We considered sampling parameter values $sample\_ct= \{1,2,8,32,64,256,4096\}$. Table~\ref{comp:repair} shows the compression ratios of the RePair-compressed text that corresponds to each sampling configuration. Yet, the plot included in the parent paper (Fig.10-left) when $80$ chars included only points for $sample\_ct= \{1,2,8,32,64\}$, and when extracting $13,000$ chars (see Figure~\ref{fig:exp.pos.extract}, or Fig.10-right in the parent paper) we tuned $sample\_ct= \{1,8,32,256,4096\}$. 

\begin{table}
	\centering
	\scriptsize	
	\begin{minipage}[]{.50\textwidth}				
	\setlength{\tabcolsep}{0.5em} 
	{\renewcommand{\arraystretch}{1.2}
		\begin{tabular}{|l|r|l|}
			\hline
			\multicolumn{3}{|c|}{\bf Non-positional indexes} \\
			\hline
			\multirow{2}{*}{\bf Method}	& \multicolumn{2}{c|}{\bf Compression ratio} \\
			\cline{2-3}
			& \multicolumn{1}{c|}{\bf Value} 	& \multicolumn{1}{c|}{\bf Parameterization} \\
			\hline
			\hline
			\vbyte\							& 4.4592\% & \multicolumn{1}{c|}{$\times$} \\\hline	
			\vbyteB\ 						& 3.3522\% & $lenBitmapDiv = 8$ \\\hline	
			\multirow{2}{*}{\vbyteCM} 		& 4.4956\% & $k=32$ \\\cline{2-3}
											& 4.7095\% & $k=4$ \\\hline
			\multirow{2}{*}{\vbyteCMB}  	& 3.3754\% & $k=32, lenBitmapDiv = 8$ \\\cline{2-3}
											& 3.4996\% & $k=4,~ lenBitmapDiv = 8$ \\\hline
			\multirow{2}{*}{\vbyteST} 		& 4.5147\% & $B=128$ \\\cline{2-3}
											& 4.8650\% & $B=16$ \\\hline
			\multirow{2}{*}{\vbyteSTB} 		& 3.3820\% & $B=128,lenBitmapDiv = 8$ \\\cline{2-3}
											& 3.5488\% & $B=16, ~ lenBitmapDiv = 8$ \\\hline		
			\rice\ 							& 3.0807\% & \multicolumn{1}{c|}{$\times$} \\\hline	
			\riceB\ 						& 3.2235\% & $lenBitmapDiv = 8$ \\\hline	
			\simplen\ 						& 0.9130\% & \multicolumn{1}{c|}{$\times$} \\\hline	
			\pfordelta\ 					& 0.9372\% & $pfdThreshold = 100$ \\\hline	
			\qmx\ 							& 4.8825\% & \multicolumn{1}{c|}{$\times$} \\\hline
			\riceRuns\ 						& 0.3247\% & \multicolumn{1}{c|}{$\times$}  \\\hline
			\vbyteLZMA\ 					& 0.2030\% & $minbcssize=10$  \\\hline		
			\multirow{4}{*}{\vbyteLzend}	& 0.1420\% & $ds=256$ \\ \cline{2-3}
											& 0.1437\% & $ds=64$ \\ \cline{2-3}
											& 0.1508\% & $ds=16$ \\ \cline{2-3}
											& 0.1790\% & $ds=4$ \\\hline		
			\repair\ 					    & 0.1040\% & \multicolumn{1}{c|}{$\times$}  \\\hline		
			\repairSkip\ 					& 0.1097\% & \multicolumn{1}{l|}{~~~~~~~~~~~~$repairBreak=4\!\times\! 10^{-7}$}  \\\hline		
			\multirow{2}{*}{\repairSkipCM}	& 0.1195\% & $k=64$, ~~$repairBreak=4\!\times\! 10^{-7}$ \\ \cline{2-3}
											& 0.1212\% & $k=1$, ~~~$repairBreak=4\!\times\! 10^{-7}$ \\ \hline 
			\multirow{1}{*}{\repairSkipST}	& 0.1279\% & $B=2^{10}$,~$repairBreak=4\!\times\! 10^{-7}$ \\ \cline{2-3}
											\hline		
			\efopt\ 						& 0.4984\% & \multicolumn{1}{c|}{$\times$} \\\hline	
			\optpfd\ 						& 0.7015\% & \multicolumn{1}{c|}{$\times$} \\\hline	
			\interpolative\ 				& 0.5573\% & \multicolumn{1}{c|}{$\times$} \\\hline	
			\varint\ 						& 5.3144\% & \multicolumn{1}{c|}{$\times$} \\\hline			
		\end{tabular}
    }
	\medskip

	\end{minipage} 
	\begin{minipage}[]{.49\textwidth}
	\setlength{\tabcolsep}{0.5em} 
	{\renewcommand{\arraystretch}{1.2}
		\begin{tabular}{|l|r|l|}
			\hline
			\multicolumn{3}{|c|}{\bf Positional indexes} \\
			\hline
			\multirow{2}{*}{\bf Method}	& \multicolumn{2}{c|}{\bf Compression ratio} \\
			\cline{2-3}
			& \multicolumn{1}{c|}{\bf Value} 	& \multicolumn{1}{c|}{\bf Parameterization} \\
			\hline
			\hline
			\vbyte\			& 35.1530\% & \multicolumn{1}{c|}{$\times$} \\\hline
			\vbyteCM		& 35.3805\% & $k=32$ \\ \cline{2-3}
							& 36.5899\% & $k=4$ \\\hline
			\vbyteST\		& 35.4935\% & $B=128$ \\ \cline{2-3}
							& 37.4281\% & $B=16$ \\\hline
			\rice\ 			& 36.7974\% & \multicolumn{1}{c|}{$\times$}	\\\hline
	
			\simplen\ 		& 36.6233\% & \multicolumn{1}{c|}{$\times$} \\\hline
			\qmx\ 			& 136.8762\% & \multicolumn{1}{c|}{$\times$} \\\hline
			\vbyteLZMA\		& 9.7539\% &  $minbcssize=10$ \\\hline
			\repairNo\ 		& 20.0641\% & \multicolumn{1}{c|}{$\times$} \\\hline
			\repairSkip\ 	& 21.3769\% & \multicolumn{1}{c|}{~~~~~~~~$repairBreak=5\times 10^{-7}$} \\\hline
			\repairSkipCM	& 20.0786\% & $k=64$, ~~$repairBreak=5\times 10^{-7}$ \\ \cline{2-3}
							& 20.8584\% & $k=1$, ~~~~$repairBreak=5\times 10^{-7}$ \\\hline
			\repairSkipST	& 21.7456\% & $B=256$, $repairBreak=5\times 10^{-7}$ \\ \hline 
			\efopt\ 		& 30.5630\% & \multicolumn{1}{c|}{$\times$} \\\hline
			\optpfd\ 		& 29.7424\% & \multicolumn{1}{c|}{$\times$} \\\hline
			\interpolative\ & 29.8820\% & \multicolumn{1}{c|}{$\times$} \\\hline
			\varint\ 		& 37.9975\% & \multicolumn{1}{c|}{$\times$} \\\hline
			\hline
			\rlcsa\			& 2.4735\% & $sample=2048$ \\ \cline{2-3}
							& 2.8686\% & $sample=1024$ \\ \cline{2-3}
							& 3.6630\% & $sample=512$ \\ \cline{2-3}
							& 5.2489\% & $sample=256$ \\ \cline{2-3}
							& 8.4048\% & $sample=128$ \\ \cline{2-3}
							& 14.6924\% & $sample=64$ \\ \cline{2-3}
							& 27.2996\% & $sample=32$ \\\hline	
			\wcsa\			& 8.8921\% & $\langle sA, sAinv,sPsi \rangle = \langle 2048,2048,2048 \rangle$ \\\cline{2-3}
							& 9.4071\% & $\langle sA, sAinv,sPsi \rangle = \langle 512,512,512 \rangle$ \\\cline{2-3}
							& 11.0718\% & $\langle sA, sAinv,sPsi \rangle = \langle 128,256,128 \rangle$ \\\cline{2-3}
							& 15.8673\% & $\langle sA, sAinv,sPsi \rangle = \langle 64,128,32 \rangle$ \\\cline{2-3}
							& 17.9458\% & $\langle sA, sAinv,sPsi \rangle = \langle 32,64,32 \rangle$ \\\cline{2-3}
							& 25.6777\% & $\langle sA, sAinv,sPsi \rangle = \langle 16,64,16 \rangle$ \\ \cline{2-3}
							& 50.8565\% & $\langle sA, sAinv,sPsi \rangle = \langle 8,8,8 \rangle$ \\\hline
			\slp\ 			& 3.2374\% & \multicolumn{1}{c|}{$\times$} \\\hline
			\wslp\ 			& 2.8962\% & \multicolumn{1}{c|}{$\times$} \\\hline
			\lzindex\		& 1.8357\%  & \multicolumn{1}{c|}{$\times$} \\\hline
			\lzendindex\	& 2.3894\%  & \multicolumn{1}{c|}{$\times$} \\\hline
		\end{tabular}
	}
	\end{minipage}
	\caption{\label{comp:indexes} Summary of compression ratios for all indexes in the framework.}
\end{table}

\begin{table}
	\centering
	\scriptsize
	\begin{tabular}{|r|l|}
		\hline
		\multicolumn{1}{|c|}{\bf Compression ratio} 	& \multicolumn{1}{c|}{\bf Parameterization} \\ 
		\hline
		\hline
		1.306\%  & sample\_ct = 1     \\\hline
		1.265\%  & sample\_ct = 2     \\\hline
		1.227\%  & sample\_ct = 8     \\\hline
		1.215\%  & sample\_ct = 32    \\\hline
		1.213\%  & sample\_ct = 64    \\\hline
		1.211\%  & sample\_ct = 256   \\\hline
		1.210\%  & sample\_ct = 4096  \\\hline	
	\end{tabular}
	\caption{\label{comp:repair} Summary of compression ratios for the RePair-compressed version of the text with different sampling values.}
\end{table}

\section{Using Self-Indexes for other types (non-document oriented) of Highly-Repetitive Collections}
\label{app:B}

While self-indexes like \wcsa\ or \wslp\ are designed to operate on words, the remaining approaches in our benchmark can 
effectively operate on an universal scenario; i.e. they are able to exploit repetitiveness underlying to any arbitrary sequence 
of chars. 

Focusing on \slp, \lzindex, and \lzend, the implementations provided in \uilib\ can be easily used to self-index any 
data collection. Scripts for indexing and searching can be reused, although the number of parameters passed to {\tt locate} 
changes. Currently, {\tt locate} requires a parameter (called {\em doc\_boundaries}) which sets the path to a file that
contains the array of offsets that mark the beginning of each document in the text. It is required to map absolute positions
to {\em (doc,offset)} pairs which indicate the position of each pattern occurrence within a document. However, this
operation is not usually needed in a general scenario, so the corresponding parameter is neither needed.

Therefore, to invoke {\tt locate} in a general scenario we have simply to remove {\em doc\_boundaries}, and the script will
deliver absolute positions of the pattern in the text.

\end{document}